\documentclass[letterpaper,10pt,nofootinbib,prd,tightenlines,twocolumn, superscriptaddress,floatfix, showpacs]{revtex4}
\pdfoutput=1
\usepackage{amsmath,amsfonts,amssymb}
\usepackage{mathrsfs}
\usepackage{graphicx}
\usepackage[english]{babel} 
\usepackage{booktabs}
\newcommand{\ra}[1]{\renewcommand{\arraystretch}{#1}}
\usepackage{appendix}
\usepackage{hyperref} 
\usepackage{multirow}

\def\eref#1{(\ref{#1})}

\begin{document}

\title{A Guide to Designing Future Ground-based CMB Experiments }
\author{W. L. K. Wu}
\email{wlwu@stanford.edu}
\affiliation{Department of Physics, Stanford University, \\382 Via Pueblo Mall, Stanford, CA 94305, USA}
\affiliation{Kavli Institute for Particle Astrophysics and Cosmology, SLAC, 
\\2575 Sand Hill Road, M/S 29, Menlo Park, CA  94025, USA}

\author{J. Errard}
\affiliation{Computational Cosmology Center - Lawrence Berkeley National Lab, \\1 Cyclotron Road, Berkeley, CA 94720, USA}
\affiliation{Department of Physics, University of California, Berkeley, CA 94720, USA}

\author{C. Dvorkin}
\affiliation{ Institute for Advanced Study, School of Natural Sciences,\\ Einstein Drive, Princeton, NJ 08540, USA}

\author{C. L. Kuo}
\affiliation{Department of Physics, Stanford University, \\382 Via Pueblo Mall, Stanford, CA 94305, USA}
\affiliation{Kavli Institute for Particle Astrophysics and Cosmology, SLAC,
\\2575 Sand Hill Road, M/S 29, Menlo Park, CA  94025, USA}

\author{A. T. Lee}
\affiliation{Department of Physics, University of California, Berkeley, CA 94720, USA}
\affiliation{Lawrence Berkeley National Laboratory, 1 Cyclotron Road, Berkeley, CA 94720, USA}

\author{P. McDonald}
\affiliation{Lawrence Berkeley National Laboratory, 1 Cyclotron Road, Berkeley, CA 94720, USA}

\author{A. Slosar}
\affiliation{Brookhaven National Laboratory, Upton, NY 11973, USA}

\author{O. Zahn}
\affiliation{Berkeley Center for Cosmological Physics, Department of Physics, University of California, and Lawrence Berkeley National Laboratory, Berkeley, CA 94720, USA}

\date{\today}

\begin{abstract}

In this follow-up work to the High Energy Physics Community Summer Study 2013 (HEP CSS 2013, a.k.a. \textsc{Snowmass}), we explore the scientific capabilities of a future Stage-IV Cosmic Microwave Background polarization experiment (CMB-S4) under various assumptions on detector count, resolution, and sky coverage.  We use the Fisher matrix technique to calculate the expected uncertainties in cosmological parameters in $\nu\Lambda$CDM that are especially relevant to the physics of fundamental interactions, including neutrino masses, effective number of relativistic species, dark-energy equation of state, dark-matter annihilation, and inflationary parameters.  To further chart the landscape of future cosmology probes, we include forecasted results from the Baryon Acoustic Oscillation (BAO) signal as measured by DESI to constrain parameters that would benefit from low redshift information. 
We find the following best 1-$\sigma$ constraints: $\sigma(M_{\nu})$ $= 15$ meV, $\sigma(N_{\rm eff})$ $= 0.0156$, Dark energy Figure of Merit = 303, $\sigma(p_{ann})$ $= 0.00588\times3\times10^{-26}$ cm$^3$/s/GeV,  $\sigma(\Omega_K)$ $= 0.00074$,  $\sigma(n_s)$ $= 0.00110$,  $\sigma(\alpha_s)$ $=  0.00145$, and  $\sigma(r)$ $= 0.00009$. We also detail the dependences of the parameter constraints on detector count, resolution, and sky coverage.
\end{abstract}


\maketitle

\section{Introduction}
In the past two decades, CMB experiments have made great strides in sensitivity improvement. For satellite experiments, there was a factor of 10 better in sensitivity from \textit{COBE} to \textit{WMAP} and from \textit{WMAP} to \textit{Planck}~\cite{2013arXiv1309.5383A}. These satellites, along with numerous ground-based and balloon-borne experiments, as listed in Ref. \cite{Lambda}, collectively moved us from setting upper limits on the temperature anisotropy ($C_\ell^{TT}$), the E- and B-mode polarization ($C_\ell^{EE}$ and $C_\ell^{BB}$) spectra to precision measurements of the temperature power spectrum (e.g.~\cite{2013ApJ...779...86S, 2013arXiv1303.5075P, 2013arXiv1301.1037D}), shrinking error bars on the E-mode polarization spectrum (e.g.~\cite{2012ApJ...760..145Q, 2013arXiv1310.1422B}), and detecting gravitational lensing in the B-mode polarization power spectrum~\cite{2013arXiv1307.5830H, 2013arXiv1312.6645P,2013arXiv1312.6646P}.

Tremendous opportunities still lie in CMB polarization. CMB polarization maps can be projected into a curl-free component (E-modes), and a divergence-free component (B-modes). Scalar and tensor perturbations seeded the temperature anisotropies in the CMB. 
While E-modes, like temperature anisotropies, could be sourced by both scalar and tensor perturbations through Thomson scattering, the only primordial source for B-modes is from tensor perturbations i.e. gravitational waves. Detecting this B-mode polarization signal, which peaks at the degree angular scale, would be a powerful probe of inflation.

B-modes at smaller angular scales are generated by gravitational lensing of E-modes by Large-Scale Structures (LSS). The conversion from E- to B-modes provides an exceptionally clean way of reconstructing the lensing potential $\phi$ and its power spectrum $C_\ell^{\phi\phi}$ \cite{2003PhRvD..67h3002O}. The latter is sensitive to the properties of the LSS, making it a probe of the late universe ($z < 10$ \cite{2006PhR...429....1L}). As a result, among other astrophysical properties, it is central to the measurements of the total neutrino mass and dark-energy equation of state. 

Different techniques have been developed to reconstruct lensing potential from CMB temperature and polarization maps \cite{2003PhRvD..67h3002O, Hirata:2003ka}. Recently, experiments have detected gravitational lensing in CMB temperature maps with high enough significance to measure the lensing potential power spectrum ~\cite{2011PhRvL.107b1301D,2012ApJ...756..142V,2013arXiv1303.5077P}. Moreover, the first reconstruction through E-to-B channel has recently been demonstrated \cite{2013arXiv1312.6646P}. 

The lensing B-mode signal also acts as a foreground to the inflationary B-mode signal. Specifically, if the tensor-to-scalar ratio $r$, which parametrizes the amplitude of the tensor perturbation, is smaller than 0.02, the level of inflationary B-modes would be lower than the lensing B-modes \cite{Kesden:2002ku,Knox:2002pe}. Algorithms have been developed to remove the lensing-induced B-modes \cite{2012JCAP...06..014S} to study the primordial signal. Therefore, measuring the lensing potential to high precision is fundamental to advancing our understanding of both the early and the late universe.

We envision the Stage-IV\footnotemark  \footnotetext{Each advance in stage refers roughly to an order of magnitude increase in number of detectors \cite{2013arXiv1309.5383A}.} Cosmic Microwave Background polarization experiment (CMB-S4) to be a high-resolution, high-sensitivity experiment for this very purpose. This paper explores how much CMB-S4 could help answer the pressing questions in High Energy Physics (HEP)  --- more broadly, physics of fundamental interactions. In what follows, we outline the main HEP topics that can be addressed by CMB polarization.\\

\textbf{Cosmic Neutrino background --} The relic neutrino number density for each species is 112 cm$^{-3}$ \cite{Fukugita}. This makes neutrinos the second most abundant particle in the universe, after photons. Although neutrinos are massless in the Standard Model, solar and atmospheric neutrino oscillation experiments have shown that they are massive. Assuming normal hierarchy, a lower limit on the total neutrino mass is $\sim58$ meV. Though minuscule in mass, their large cosmic abundance enables us to observe their cumulative gravitational effect on structure formation: neutrinos suppress structure growth below scales defined by their free streaming distances. As a result, we can measure the total neutrino mass by observing the matter power spectrum -- which CMB lensing is particularly sensitive to~\cite{2003PhRvL..91x1301K}. Another way to study the cosmic neutrino background is to measure the extra number of relativistic species $N_{\text{eff}}$ aside from photons at recombination. In the Standard Model, no other particles except neutrinos contribute as extra relativistic species. In this scenario, $N_{\text{eff}}$ is predicted to be 3.046. Deviation from this value hints at new physics. The CMB is uniquely sensitive to the photon-baryon-dark matter interaction at recombination. The current constraints on $N_{\text{eff}}$ are  $3.30^{+0.54}_{-0.51}$ (at the $95\%$ C.L.) from \textit{Planck} + \textit{WMAP} polarization + small scale CMB data + BAO measurements, and the current limit on total neutrino mass is $< 0.23\, \text{eV}$ (at the $95\%$ C.L.) from \textit{Planck} data + BAO data \cite{2013arXiv1303.5076P}. \\

\textbf{Dark Energy --} Since the discovery of dark energy \cite{1998AJ....116.1009R,1999ApJ...517..565P}, its contribution to the energy density of the universe $\Omega_{\Lambda}$ has been determined to percent level over the past 15 years.  However, its fundamental nature still remains mysterious. The three broad possibilities are: (1) vacuum energy manifesting as a cosmological constant; (2) a spatially homogeneous dynamical field; (3) modified gravity on cosmological scales.  These hypotheses can be tested experimentally by measuring the dark-energy equation of state $w$ through the clustering of LSS as a function of redshift. So far, measurements are consistent with a cosmological constant.  Over the next decade, photometric and spectroscopic surveys such as LSST, Euclid, and DESI will take these tests to the next level of precision, opening up opportunities for new discoveries \cite{Albrecht2012}.  CMB lensing is highly complementary to these surveys, because it provides high-redshift information in the linear regime, cf.~\cite{2001PhR...340..291B}.  It is especially valuable for searches of early evolution of $w$ when $z>2$, which would not be possible with galaxy surveys. In addition, CMB lensing can help optical lensing surveys calibrate their shear bias and boost constraints on cosmological parameters, \cite{2013arXiv1311.0905P, 2013arXiv1311.2338D, 2013arXiv1311.6200H}. The current constraints of $w_0$ and $w_a$ in the $w = w_0+w_a(1-a)$ model are: $w_0 = -1.04^{+0.72}_{-0.69}$ and $w_a < 1.32$  (\textit{Planck}+\textit{WMAP} polarization + BAO data), cf.~\citet{2013arXiv1303.5076P}. \\

\textbf{Dark-matter annihilation--} The gravitational properties of dark matter (DM) have been well constrained by data from a plethora of measurements, e.g. weak~\cite{2003ARA&A..41..645R} and strong~\cite{1998ApJ...498L.107T} lensing, multi-wavelength studies of the Bullet Cluster \cite{2006ApJ...648L.109C}, distant supernovae~\cite{1998AJ....116.1009R,1999ApJ...517..565P}, and the CMB~\cite{2013arXiv1303.5076P}. However, the particle nature of dark matter remains unknown. There are several types of experiments aiming to understand dark matter interactions: (1) direct detection experiments, which rely on nuclear recoil signature of DM-Standard Model particle scattering, (2) indirect detection experiments, which are sensitive to the products of DM decay or annihilation, and (3) collider experiments. For a DM theory and detection review, see Ref. \cite{2010ARA&A..48..495F}.

An alternative observable is the CMB two-point correlation function. When DM annihilates, heat is transferred to the photon-baryon fluid, atoms are ionized and/or excited~\cite{2004PhRvD..70d3502C}. \citet{2013PhRvD..87l3513S} performed a detailed study regarding the energy deposition history into the photon-baryon fluid. 

Dark-matter annihilation also leads to growing ionization fraction perturbations and amplified small-scale cosmological perturbations, leaving an imprint on the CMB bispectrum \cite{2013PhRvD..87j3522D}. CMB temperature and polarization spectra can constrain the parameter $p_{ann} \equiv  f \langle \sigma v \rangle\,/\,m_{DM}$, where $f$ is the fraction of energy deposited into the plasma, $\langle \sigma v \rangle$ is velocity-weighted cross section, and $m_{DM}$ is the mass of the DM particle. Current constraints coming from \textit{WMAP} $9$-year data, \textit{Planck}, ACT, SPT,  BAO, HST and SN data exclude DM masses below $26$ GeV at the 2$\sigma$ level, assuming that all the energy is deposited in the plasma \cite{2013arXiv1310.3815M}. We show in this paper that CMB-S4 will tighten these constraints by a factor of $10$.\\

\textbf{Inflation --} A detection of degree-scale B-mode polarization would help us learn about the physics of inflation. In particular, the amplitude of tensor perturbations is directly related to the energy scale of inflation. In addition, whether $r$ is above or below $\sim$0.01 determines whether the inflaton field range is sub- or super-Planckian in a broad class of inflationary models. Beyond B-modes, small angular scale E-mode polarization, which is less contaminated by foregrounds than the temperature maps, can test the following predictions from slow-roll inflation: (1) the nearly scalar invariant primordial power spectrum where $n_s$ is close to but not exactly 1, (2) the small running of $n_s$, and (3) the almost flat mean spatial curvature, $\Omega_K  \approx 10^{-4}$. The current tightest constraint for $n_s$ is $0.9603 \pm 0.0073$ at the $68\%$ C.L., for $dn_s/d\log k$ is $-0.014^{+0.016}_{-0.017}$ at the $95\%$ C.L., and for $\Omega_K$ is $-0.0005^{+0.0065}_{-0.0066}$ at the $95\%$ C.L. \cite{2013arXiv1303.5076P}. \\

We show in this paper, given the range of possible experimental configurations (survey coverage, depth, resolution), how sensitive future ground-based experiments like CMB-S4 will be in probing these areas of new physics. We deliberately extend the experimental configuration space stated in the Snowmass study for CMB-S4 so that we understand the steepness of the improvement in parameter constraints as a function of experimental setups. Furthermore, we detail the forecast methods which are lacking in the Snowmass report. 

The paper is organized as follows. First, in Section~\ref{sec:methods}, we describe the methodology and the fiducial cosmology.  Sections ~\ref{sec:neutrinos} to \ref{sec:Inflation} are assigned to the aforementioned HEP topics and parameters of interests:
\begin{itemize}
\item Section~\ref{sec:neutrinos}: total neutrino mass and extra relativistic species -- $M_{\nu}$ and  $N_{\rm{eff}}$
\item Section~\ref{sec:DarkEnergy}: dark-energy equation of state -- $w_0$ and $w_a$
\item Section~\ref{sec:DarkMatter}: dark-matter annihilation -- $p_{ann}$
\item Section~\ref{sec:Inflation}: inflation -- $\Omega_K$, $n_s$, $\alpha_s$ (running of $n_s$), $r$, and $n_t$
\end{itemize}

For each section above, we present a brief overview of the underlying physical phenomena and their effects on the CMB power spectra,  and present and discuss the results of the forecast. For the tensor-to-scalar ratio $r$, we have a more thorough method section outlined in Section~\ref{sssec:r-method}.  Finally, we conclude and summarize the main results in Section~\ref{sec:conclusion}.

\section{Methods and Assumptions}
\label{sec:methods}

In this section, we describe the Fisher matrix formalism we use in this work to forecast cosmological constraints, spanning over a large grid of experimental inputs. 
We also lay out the fiducial cosmology that we assume in this work, unless otherwise stated.

\begin{center}
\begin{table}[htbp]\centering \footnotesize
\ra{1.3}
\begin{tabular}{@{}r|p{0.1cm}p{0.875cm}p{0.875cm}p{0.875cm}p{0.875cm}p{0.875cm}p{0.75cm}p{0.75cm}@{}}\hline
$N_{\rm det}$ $\backslash$ f$_{\text{sky}}$ && 0.0125& 0.025 & 0.05 & 0.125 & 0.25 & 0.5 & 0.75 \\ \hline
10,000 \ \  && 0.75 & 1.06 &1.50 & 2.37 & 3.34 & 4.73 & 5.79  \\
20,000 \ \ && 0.53 & 0.75 & 1.06 & 1.67 & 2.37 & 3.34  & 4.10 \\
50,000 \ \ && 0.33 & 0.47 & 0.67 & 1.06 & 1.50 & 2.12 & 2.59  \\
100,000 \ \  && 0.24 & 0.47 & 0.47 & 0.75 & 1.06 & 1.50 & 1.83 \\
200,000 \ \   && 0.17 & 0.33 & 0.33  & 0.53 & 0.75 & 1.06 & 1.30 \\
500,000 \ \  && 0.11 & 0.21 & 0.21 & 0.33 & 0.47 & 0.67 & 0.82 \\
1,000,000 \ \   && 0.075 & 0.15 & 0.15 & 0.24 & 0.33 & 0.47 & 0.58 \\
\hline
\end{tabular}
\caption{Table of experimental sensitivities $s$ (multiply by $\sqrt{2}$ for polarization) given detector count (left column) and sky coverage (top row), in $\mu$K-arcmin}
\label{table:sensifsky}
\end{table}
\end{center}

\subsection{Experiment Input Grid}
\label{ssec:exp_setup}

The exact configuration of CMB-S4 is still being studied.
A few examples were given in the \textsc{Snowmass} CF5 Neutrinos report \cite{2013arXiv1309.5383A}. 
Here we extend the experiment input grid (sensitivity, beam size, sky coverage $f_{sky}$) to understand how constraints for each parameter vary across the experiment design space.

The experimental sensitivity in $\mu$K-arcmin is derived from the combination of total number of detectors and the observed fraction of sky. Table \ref{table:sensifsky} presents the sensitivity of an experiment given a total number of detectors $N_{\rm det}$ and a sky coverage $f_{sky}$. We consider $N_{\rm det}$ ranging from $10^4$ to $10^6$ and $f_{sky}$ from $1\%$ to $75 \%$. We assume a Noise-Equivalent Temperature (NET) of $350\, \mu$K.$\sqrt{s}$ per detector (achieved by current generation experiments), and a yield (Y) of $25\%$ that combines both the focal plane yield and the operation efficiency over an observing period of 5 years ($\Delta T$). To get from NET per detector to the sensitivity $s$ of an experiment, we use
\begin{eqnarray}
\centering
	s\ [\;\mu {\rm K.arcmin}\;] \equiv \frac{ {\rm NET}\ [\; \mu{\rm K.}\sqrt{s}\;] \times \sqrt{f_{sky} \ [\;{\rm arcmin}^2\;] }}{ \sqrt {N_{\rm det} \times Y \times \Delta T\ [\;{\rm s }\;]}}.
	\label{eq:sensitivity_definition}
\end{eqnarray}
For each combination of $N_{\rm det}$ and $f_{sky}$, we cover the following beam sizes: $1'$, $2'$, $3'$, $4'$ -- and in some cases up to $6'$ and $8'$.

\subsection{Fisher matrix}
The Fisher matrix formalism is a simple way to forecast how well future CMB experiments can constrain parameters in the cosmological parameter space of interest. 
By definition, Fisher matrix assumes a Gaussian likelihood and it may not represent the true likelihood distribution in some cases \cite{2012JCAP...09..009W}. This approach has been scrutinized and other methods for forecast have been proposed \cite{2006JCAP...10..013P, 2012JCAP...09..009W}. However, for a large grid of experimental inputs, it is computationally efficient to arrive at constraints through Fisher information. It is also useful for comparison with previous work, e.g. \cite{2006astro.ph..9591A, 2013arXiv1308.4164F}, and within the grid. 

Assuming the likelihood function $\mathcal L$ of the parameters $\boldsymbol{\theta}$ given data $\mathbf{d}$ to be Gaussian, it can be written as
\begin{equation}
\mathcal{L}(\boldsymbol{\theta}| \boldsymbol{d}) \propto \frac{1}{\sqrt{| \mathbf{C}(\boldsymbol{\theta})|}} \exp \left( -\frac{1}{2} \boldsymbol{d^{\dagger}}\large\left[ \mathbf{C}(\boldsymbol{\theta})\large\right]^{-1} \boldsymbol{d}\right),
\label{eq:likelihood_def}
\end{equation}
where $\mathbf{C}$ is the covariance matrix of the modeled data. In our case,  $\boldsymbol{d}= \{a_{\ell m}^T, a_{\ell m}^E, a_{\ell m}^d \}$ for temperature, E-mode polarization, and lensing-induced deflection. The vector $\boldsymbol{\theta}$ contains the cosmology model described in Section~\ref{ssec:fidcosmo}.

The Fisher matrix is constructed from the curvature of the likelihood function at the fiducial values of the parameters:
\begin{equation}
	\centering
		F_{ij} \equiv - \left\langle\frac{\partial^2 \log \mathcal{L}}{\partial \theta_i \partial \theta_j} \bigg|_{\boldsymbol{\theta} = \boldsymbol{\theta_0}}\right\rangle
	\label{eq:Fij_def}
\end{equation}
where $\boldsymbol{\theta_0}$ is a vector with the fiducial parameters, which are chosen to maximize the likelihood. 

From Eqs.~\eref{eq:likelihood_def} and \eref{eq:Fij_def}, we can write the Fisher components $F_{ij}$ for CMB spectra as
\begin{equation}
 F_{ij} = \sum_\ell \frac{2\ell+1}{2} f_{sky} {\rm Tr} \left(  \boldsymbol{C}^{-1}_\ell( \theta) \frac{\partial \boldsymbol{C}_\ell}{\partial \theta_i} \boldsymbol{C}^{-1}_\ell( \theta) \frac{\partial \boldsymbol{C}_\ell}{\partial \theta_j}  \right)
 \end{equation}
 where $\ell$ is the multipole of the angular power spectra and
 \begin{eqnarray}
 	\centering
		\mathbf{C}_\ell \equiv \left( \begin{array}{ccc}C_\ell^{TT} + N_\ell^{TT} & C_\ell^{TE} & C_\ell^{Td} \\ C_\ell^{TE} & C_\ell^{EE} + N_\ell^{EE} & 0 \\ C_\ell^{Td} & 0 & C_\ell^{dd} + N_\ell^{dd}\end{array}\right).
	\label{eq:cov_definition}
\end{eqnarray}
Similar to Ref. \cite{2009AIPC.1141..121S}, we do not consider a B-mode power spectrum signal in this covariance matrix since we assume that the fields are unlensed and that there are no primordial B-modes. We set $C_{\ell}^{Ed}$ to 0 because its effect is small ($< 1\%$ improvement on constraints in $M_{\nu}$ for most cases). $C_{\ell}^{dd}$ is related to the lensing power spectrum $C_{\ell}^{\phi\phi}$ by $C_{\ell}^{dd} = \ell(\ell+1)C_{\ell}^{\phi\phi}$.

$1$-$\sigma$ uncertainties of each parameter are obtained by marginalizing over the other parameters. For a parameter $\theta_i$, the marginalized error is given by
\begin{equation}
\sigma_i \equiv \sigma (\theta_i) = \sqrt{(\mathbf{ F^{-1}})_{ii}}
\end{equation}
The power spectra include Gaussian noise $N^{XX'}_\ell$ defined as
 \begin{equation}
 	\centering
		N^{ XX' }_\ell = s^{\, 2} \exp \left(\ell(\ell+1) \frac{\theta^{\ 2}_{\textsc{fwhm}}}{8\log2}\right)
	\label{eq:beamnoise}
\end{equation}
where $s$ is the total intensity instrumental noise in $\mu$K-radians, defined in Eq.~\eref{eq:sensitivity_definition}, and $\theta^{\ 2}_\textsc{fwhm}$ is the full-width half-maximum beam size in radians. Note that  $s \rightarrow s\times \sqrt{2}$ in the case of $ XX' = \{ EE, BB \}$.

\begin{figure}[htbp]
               \includegraphics[width=\linewidth]{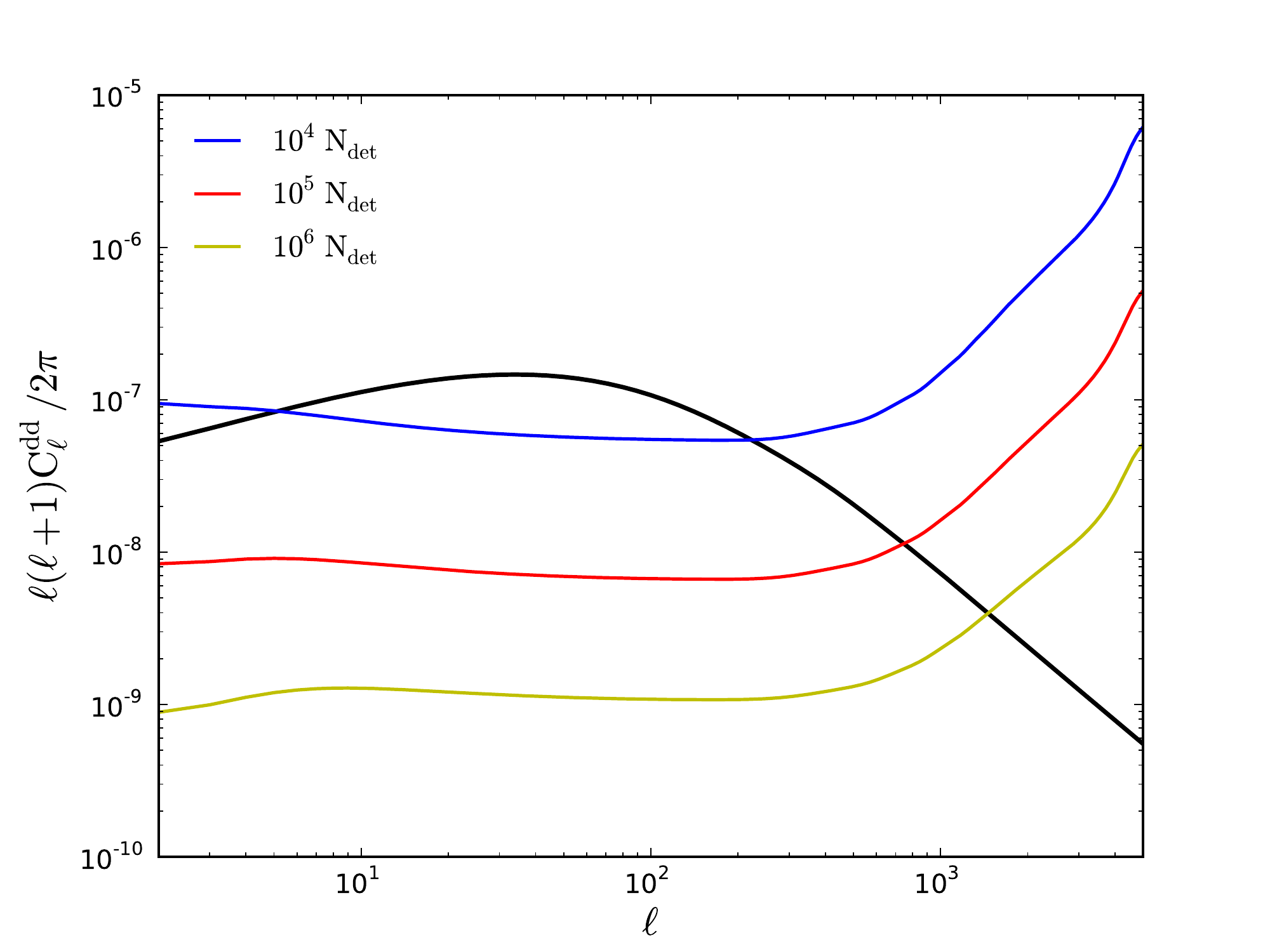}
       \caption{$N_{\ell}^{dd}$ for three $N_{\rm det}$ at $f_{sky}=0.75$ and $1'$ beam size. The deflection angle spectrum $C_{\ell}^{dd}$ is shown in black as reference. $C_{\ell}^{dd}$ is related to the lensing power spectrum $C_{\ell}^{\phi\phi}$ by $C_{\ell}^{dd} = \ell(\ell+1)C_{\ell}^{\phi\phi}$.}
       \label{fig:Nlddnoise}
\end{figure}
For $XX' = dd$, we use the iterative method outlined in ~\citet{2012JCAP...06..014S}, based on \citet{Hirata:2003ka} and \citet{2003PhRvD..67h3002O}. We estimate $N^{dd}_\ell$ from lensing reconstruction using $C^{EE}_\ell$ and $C^{BB}_\ell$ only. Fig.~\ref{fig:Nlddnoise} shows the $N_{\ell}^{dd}$ level for three $N_{\rm det}$ at $f_{sky}= 0.75$ and $1'$ beam.
We use the EB estimator because it has the lowest noise for the range of sensitivities relevant for CMB-S4 ($10^5 \lesssim N_{ \rm det} \lesssim 10^6$). For reference, the range of $\ell(\ell+1) N^{dd}_\ell /2\pi$ noise floor for $10^5 < N_{ \rm det} < 10^6$ from the EB estimator is between $10^{-8}$ and $10^{-10}$, while \textit{Planck}'s $\ell(\ell+1) N^{dd}_\ell /2\pi$ noise floor is $\sim 2 \times 10^{-7}$ from the $143$ GHz and $217$ GHz channels \cite{2013arXiv1303.5077P}. We note that, for a couple of cases with $\mathcal{O} \left( 10^4 \right)$ detectors, the TT estimator gives lower reconstruction noise for sensitivity $s \gtrsim 4\;\mu$K-arcmin and the beam sizes we consider in this paper. To demonstrate the effect of using the TT estimator instead of the EB estimator, the constraint for $M_{\nu}$, which relies most heavily on lensing information, improves from $49$~meV to $47$~meV for the case with $10^4$ detectors, $1'$ beam, and $f_{sky} = 0.75$. 
One can also combine the estimators from all reconstruction channels to get a minimum variance $N^{dd}_\ell$ to further improve the constraints if the systematics and foregrounds from the other channels are well understood. The iterative delensing method improves on the quadratic method once the instrumental noise is below the lensing B-mode level ($s\sim 3.5\mu $K-arcmin, i.e. polarization noise $\sim 5\mu $K-arcmin) \cite{2004PhRvD..69d3005S}.

\subsection{Low-redshift information}
For parameter constraints that could benefit from low-redshift information, we use the BAO signal forecasted from the Dark Energy Spectroscopic Instrument (DESI) and put 1\% priors on $H_0$, which is achievable in the next decade. 

DESI will measure the spectra of approximately 24 million objects in the 
redshift range $0<z<1.9$ (we do not include the Ly-$\alpha$ forest at 
$z>1.9$ for simplicity). 
We limit ourselves to use only the DESI BAO signal, because it is the 
most systematically robust and time-tested measurement -- constraints will be
limited by statistical errors even at the high level of precision of DESI 
\cite{2013arXiv1308.0847L}. 

While constraints from DESI broadband power spectrum and redshift space 
distortion measurements will be competitive for many of the models discussed 
in this work, 
we do not include them because modeling uncertainties related to galaxy bias 
make projections for them
less certain and we want to focus on the effect of CMB lensing.
This way, we will have strong independent measurements from both CMB lensing 
and optical surveys for these parameters in the coming decade.

Because the BAO signal is completely independent from CMB measurements including the CMB lensing signal, we add the two Fisher matrices,
\begin{equation}
F_{total} = F_{\rm CMB} + F_{\rm BAO},
\end{equation}
when BAO is included in the forecast. The independence holds even when the observing patch overlaps, as one cannot measure the BAO signal from CMB lensing maps.
Details of DESI modeling can be found in \citet{2013arXiv1308.4164F}. On the practical level, when we use the matrices from that paper, we transformed them to the parameter space of this paper.

To add a 1\% $H_0$ prior to the Fisher matrix, we set
\begin{equation}
F_{H_0 H_0} \rightarrow F_{H_0 H_0} + \frac{1}{(1\% \times H_{0,fid})^2}, 
\end{equation}
where $F_{H_0 H_0} $ is the diagonal entry for $H_0$ in the Fisher matrix, and $H_{0,fid}$ is the fiducial value of $H_0$.

\subsection{Fiducial cosmology}
\label{ssec:fidcosmo}

The fiducial cosmology is a flat $\nu \Lambda$CDM universe, with parameter values from Table 2 of \textit{Planck} best fit value \cite{2013arXiv1303.5076P}, i.e. $\Omega_c h^2 =  0.12029$,  $\Omega_b h^2  = 0.022068$, $A_s = 2.215\times10^{-9}$ at $k_0 = 0.05\ {\rm Mpc}^{-1}$, $n_s = 0.9624$, $\tau = 0.0925$, and $H_0 = 67.11$ km/s/Mpc. The fiducial $\Omega_{\nu} h^2$ is set at 0.0009, which corresponds to $M_{\nu}$  $\simeq$ 85\ meV.

Unless otherwise stated, for all extensions of the minimal model, we vary all of the above parameters (including neutrino mass) and marginalize over them to obtain the parameter constraints.
The parameters investigated in this work are:
\begin{itemize}
\item $M_{\nu}$ and $N_{\rm eff}$, see section~\ref{sec:neutrinos}; 
\item $w_0$ and $w_a$, see section~\ref{sec:DarkEnergy}; 
\item $p_{ann}$ in Section~\ref{sec:DarkMatter};
\item $\Omega_K$, $n_s$, $\alpha_s$, $r$, and $n_t$, see section~\ref{sec:Inflation}.
\end{itemize}
The $\ell_{max}$ used throughout this paper is 5000, in all the auto- and cross- spectra. Note that high-$\ell$ polarized radio sources are expected to contribute to the polarization power beyond $\ell \sim 5000$~\cite{2009AIPC.1141..121S}. 
Upper limits on the mean-squared polarization fraction of radio sources and dusty galaxies have been placed by previous studies \cite{2011MNRAS.413..132B, 2007MNRAS.374..409S}. 
Given knowledge of their mm-wave flux density distribution \cite{2010ApJ...719..763V, 2011ApJ...731..100M, 2011A&A...536A...7P}, we can identify and mask the extragalactic polarized sources with small data loss. 
We can avoid polarized galactic foregrounds by scanning patches that are outside of the Milky Way. 
Foregrounds in the temperature power spectrum, however, start to dominate over the signal around $\ell=3000$ \cite{2012ApJ...755...70R, 2013JCAP...07..025D, 2013arXiv1303.3535C, 2013arXiv1309.0382P}. The different sources of high-$\ell$ foregrounds in the temperature are thermal and kinetic Sunyaev-Zel'dovich (SZ) effects, radio galaxies, and cosmic infrared background. Progress is being made towards an understanding of these foregrounds, as many studies are currently underway~\cite{2012ApJ...755...70R, 2013JCAP...07..025D, 2013arXiv1303.3535C, 2013arXiv1309.0382P}. For parameters that rely on high-$\ell$ information, we also estimate parameters constraints with $\ell = 3000$ cut in the temperature spectrum.

\section{Cosmic Neutrino Background}
\label{sec:neutrinos}
\subsection{Overview}

The standard hot Big Bang model predicts a relic sea of neutrinos -- the cosmic neutrino background, whose density is the second highest of all species. Neutrinos decoupled from the primordial plasma at a temperature of $T \sim 1$ MeV \cite{2006PhR...429..307L}, but maintained temperature equilibrium with photons as the universe expanded. Once the temperature of the photons dropped below the mass of the electrons, electron-positron annihilation transferred heat to photons. Using entropy conservation, and assuming neutrinos had completely decoupled from the plasma by that time, we can relate the temperatures of neutrinos and photons by 
\begin{eqnarray}
	T_{\nu} = \left( \frac{4}{11} \right)^{1/3} T_{\gamma} 
	\label{eq:tnu_propto_tgamma}
\end{eqnarray}
where the factor of $4/11$ comes from the effective degrees of freedom of positrons, electrons, and photons before and after electron-positron annihilation is complete \cite{2002astro.ph..8186S}.

To account for the radiation-like behavior neutrinos that have in the primordial plasma, it is conventional to parametrize the relativistic energy density $\rho_R$ as a function of photon energy density  $\rho_{\gamma}$ and $N_{\text{eff}}$, the effective number of relativistic species, as
\begin{equation}
\rho_R = \left( 1 + N_{\text{eff}} \frac{7}{8} \left( \frac{4}{11} \right)^{4/3} \right) \rho_{\gamma}
\end{equation}
where the factor of $7/8$ accounts for the fermionic degrees of freedom of neutrinos. Given three neutrinos in the Standard Model of particle physics,  one would naively expect $N_{\rm{eff}}=3$. However, QED corrections in the primordial plasma and spectral distortion due to electron-positron annihilation raise the value of $N_{\rm{eff}}$ to 3.046 \cite{2005NuPhB.729..221M}. Because of the generic parameterization, $N_{\rm{eff}}$ encapsulates radiation-like behavior from any relativistic species as well. Sterile neutrinos, a candidate for non-standard radiation content, could change $N_{\rm{eff}}$ based on their masses and mixing angles with active neutrinos~\cite{2011arXiv1110.6479F}. Therefore, tight constraints on $N_{\rm{eff}}$ from the CMB and large-scale structures, along with Big Bang nucleosynthesis (BBN), would be essential to either confirm the Standard Cosmological Model or discover new physics relating to extra relativistic species.

Total neutrino mass, $M_{\nu}$, contributes to the critical density of the universe as \cite{2005NuPhB.729..221M}
\begin{eqnarray}
	\centering
		\Omega_{\nu}h^2 \simeq \frac{M_{\nu}}{94\, \text{eV}}.
\end{eqnarray}

\begin{figure}[htbp]
		\includegraphics[width=\linewidth]{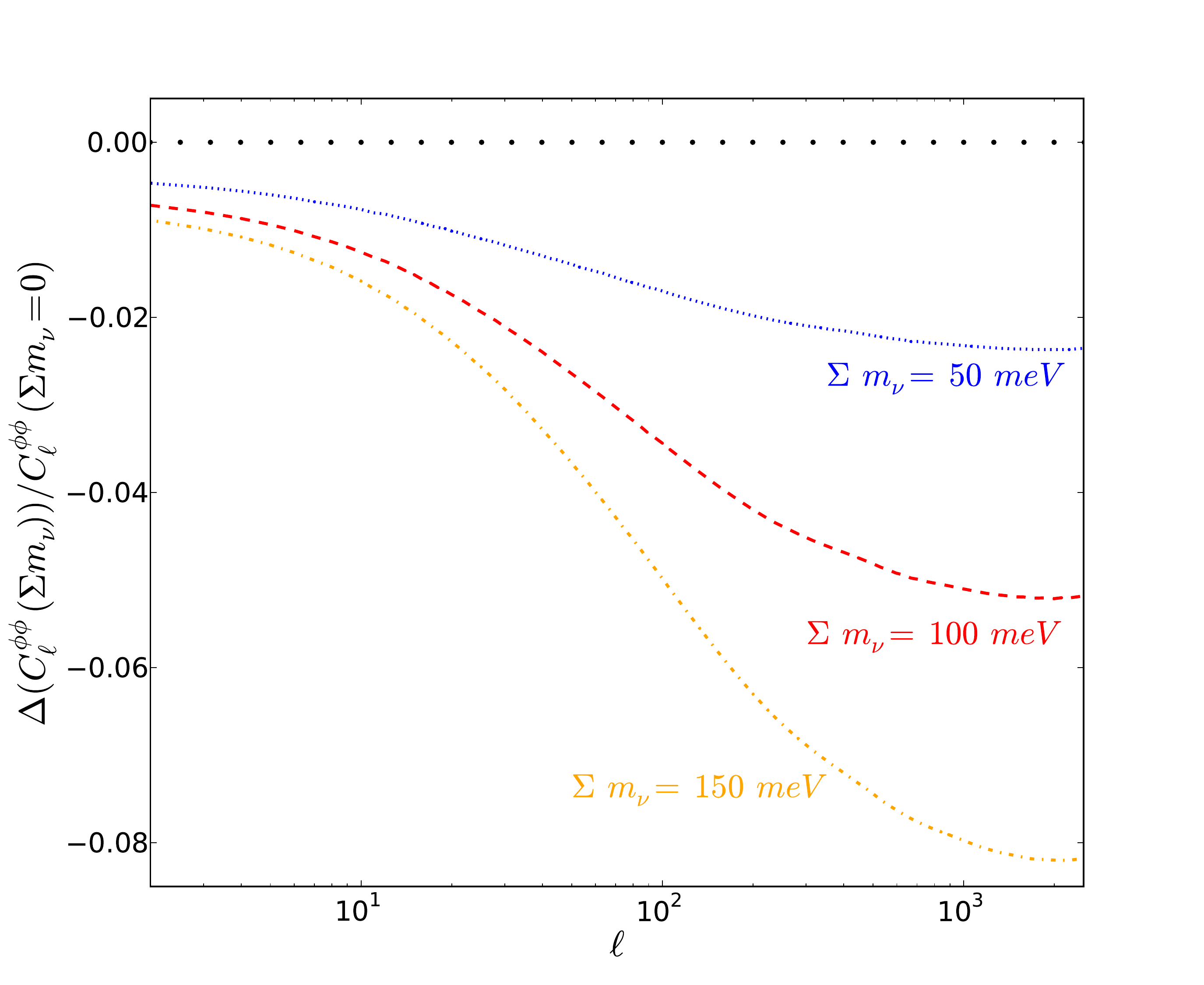}
	\caption{ Ratio of the lensing potential power spectrum for different total neutrino masses to a massless neutrino case: the heavier the neutrinos, the more suppressed the potential is.}
	\label{fig:numassCdd}
\end{figure}

\textit{Main Effects of $M_{\nu}$ and $N_{\text{eff}}$ on CMB spectra -- }
\begin{enumerate}
  \item \textit{Suppression on small angular scale lensing power spectrum for non-zero $M_{\nu}$}. Neutrinos' large thermal velocities allow them to free stream on scales smaller than $\sim (T_{\nu}/m_{\nu}) \times (1/H)$ ~\cite{2011APh....35..177A}, where $m_{\nu}$ is the mass of an individual neutrino and $H$ is the Hubble expansion rate, instead of falling into gravitational wells. As a result, structure formation below this scale is suppressed \cite{2003PhRvL..91x1301K}.  Fig.~\ref{fig:numassCdd} shows the effect of neutrino free streaming on the lensing potential power spectrum $C_\ell^{\phi\phi}$ for various total neutrino masses. 
  \item  \textit{Increasing $N_{\text{eff}}$ increases Silk damping.} The main effects of $N_{\text{eff}}$ on the CMB temperature and E-mode polarization power spectra is increased Silk damping at small scales~\cite{2013PhRvD..87h3008H}.
 \end{enumerate}

\begin{figure*}[!htbp]
\includegraphics[width=\linewidth]{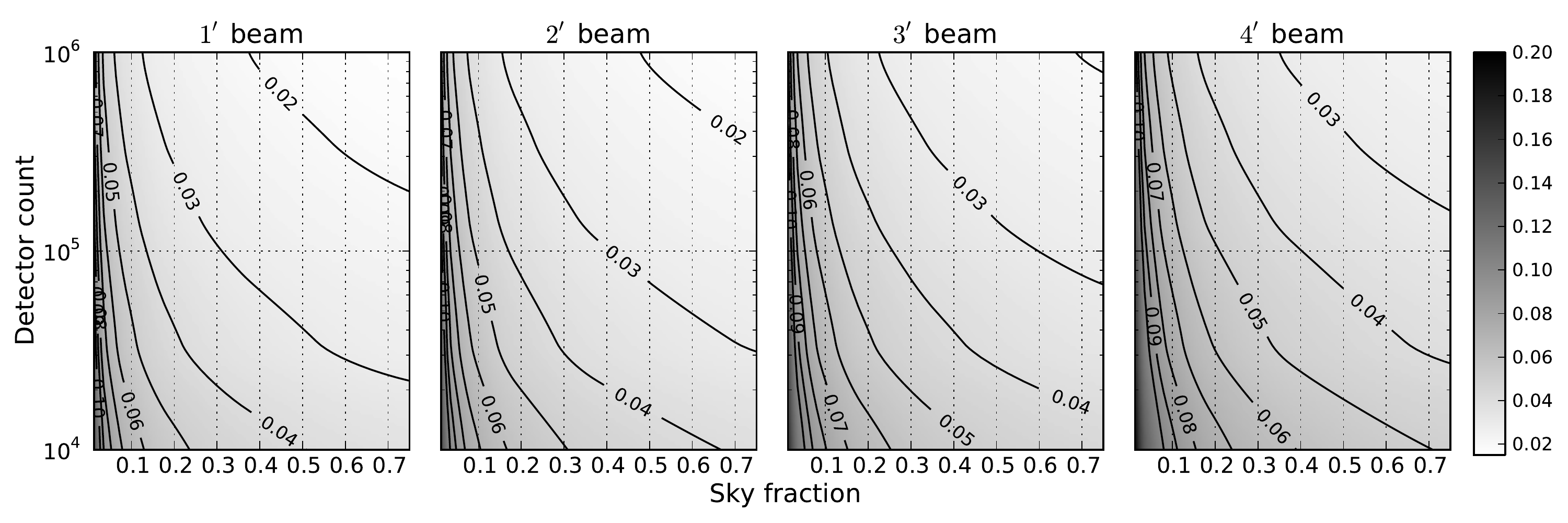}
\caption{Constraints for $\sigma(N_{\rm{eff}})$ as a function of the number of detectors and observing sky fraction for $1'$ to $4'$ beam sizes.}
\label{fig:Neff_constraints_contour}
\end{figure*}

The current constraint on $N_{\text{eff}}$ is $3.30^{+0.54}_{-0.51}$ (95\% C.L.) from \textit{Planck} + \textit{WMAP} polarization + small scale CMB + BAO data, and the current limit on total neutrino mass is $M_\nu < 0.23\, \text{eV}$ (95\% C.L.) from \textit{Planck} + BAO data \cite{2013arXiv1303.5076P}.  From solar and atmospheric neutrino oscillation experiments, we know the mass differences between each mass eigenstate satisfy:
\begin{eqnarray}
	\centering
		\begin{aligned}
		m_2^2 - m_1^2 &=& 7.54^{+0.26}_{-0.22} \times 10^{-5}\ \text{eV}^2\\
		m_3^2 - \frac{m_1^2+m_2^2}{2} &=& \pm 2.43^{+0.06}_{-0.10} \times 10^{-3}\ \text{eV}^2.
     		\end{aligned}
\end{eqnarray}
From these equalities, the lower bound for the sum of neutrino masses is $\sim 58$ meV for the normal hierarchy, where the smallest mass is set to zero, and $\gtrsim 100$ meV for the inverted hierarchy, in which case $m_3$ is set to have the smallest mass~\cite{2006PhR...429..307L}. 

The next qualitatively interesting result for neutrino mass through cosmological probes will be to detect the total neutrino mass with significance, and potentially distinguish normal from inverted hierarchies if the smallest mass is $< 100$ meV. In addition, precise measurement of $N_{\text{eff}}$ would determine whether the three active neutrinos scenario accurately describes the thermal content of the early universe. An order of magnitude improvement from current constraints are likely to be needed to achieve this goal.

Experiments like Katrin\footnotemark  \footnotetext{http://www.katrin.kit.edu/} directly probe the effective electron neutrino mass. Coupled with neutrino mixing angle measurements like No$\nu$a\footnotemark  \footnotetext{http://www-nova.fnal.gov/}, T2K\footnotemark \footnotetext{http://t2k-experiment.org/} or Double Chooz\footnotemark \footnotetext{http://doublechooz.in2p3.fr/}, one can pin down the masses of each neutrino species. Agreement will confirm our understanding of cosmology and particle physics, and disagreement will initiate search for new physics.

\subsection{Results and discussion}
\subsubsection{$N_{\text{eff}}$ forecast}
\label{ssec:neff}

Table ~\ref{table:Neff} presents a sample of $N_{\rm eff}$ constraints given different beam sizes, sky coverage, and detector counts in an experiment. The best constraints we get in this grid is $\sigma(N_{\rm eff})=0.016$ from the $10^6$ detectors, $1'$ beam, $f_{sky}=0.75$ case (equivalent sensitivity is $0.58\,\mu$K-arcmin), which distinguishes $3.046$ from 3 at the $3$-$\sigma$ level. 
We observe that increasing $N_{\rm det}$ from $10^4$ to $10^5$ improves constraints in $N_{\rm eff}$ by about 35\%, while increasing $N_{\rm det}$ from $10^5$ to $10^6$ improves constraints in $N_{\rm eff}$ by about 30\%. 
Increasing $f_{sky}$ also improves the constraints even though the sensitivity of the experiment decreases. 
The constraints improve by about 10\% for each arc-minute decrease in beam size. Out of all the parameters studied in this work, $N_{\rm eff}$ improves most with decreasing beam size. 
This is partly due to the high constraining power on $N_{\rm eff}$ from high $\ell$ multipoles in the $C_\ell^{TT}$ and $C_\ell^{EE}$ spectra which are sensitive to effects from Silk damping. 

\begin{table}[ht]\centering
\ra{1.3}
\begin{tabular}{@{}lccccl@{}}\hline
beams & $1'$ & $2'$ & $3'$ & $4'$ \\ 
\hline
$\mathbf{10^4}$ \textbf{detectors} \\
$f_{sky} = 0.25$ & 4.91 & 5.34 & 6.02 & 6.90 \\
$f_{sky} = 0.50$ & 4.02 & 4.36 & 4.88 &  5.57 \\
$f_{sky} = 0.75$ & 3.60 & 3.88 & 4.33 &  4.93 \\
$\mathbf{10^5}$ \textbf{detectors} \\
$f_{sky} = 0.25$ & 3.24 & 3.54 & 4.04 & 4.71 \\
$f_{sky} = 0.50$ & 2.57 & 2.81 & 3.19 & 3.72\\
$f_{sky} = 0.75$ & 2.25 & 2.46 & 2.79 & 3.24 \\
$\mathbf{10^6}$ \textbf{detectors} \\
$f_{sky} = 0.25$ & 2.32 & 2.53 & 2.89 & 3.41 \\
$f_{sky} = 0.50$ & 1.80 & 1.97 & 2.25 & 2.64 \\
$f_{sky} = 0.75$ & 1.56 &1.70 & 1.94 & 2.28 \\
\hline
\end{tabular}
\caption{The constraints on $\sigma(N_{\rm eff})$ derived from CMB in units of $10^{-2}$ for various combinations of beam sizes, sky coverage, and detector counts.}
\label{table:Neff}
\end{table}

We know that for multipoles $\ell>3000$, the temperature power spectrum is also contaminated by foregrounds. For this reason, we run the forecast up to $\ell=3000$ in $C_\ell^{TT}$ while keeping the other spectra the same. In this case, we get $N_{\rm{eff}}$ 1-$\sigma$ constraint at 0.021 with $10^6$ detectors, $f_{sky}=0.75$, $1'$ beam. The set of experiments that could constrain $N_{\rm{eff}}$ to 0.025 shrinks to \{$N_{\rm det}$, beam, $f_{sky}$\} = \{$10^6$, 0.75, $1'$\},   \{$5 \times 10^5$, 0.75, $1'$\},  \{$2 \times 10^5$, 0.75, $1'$\}, \{$10^6$, 0.5, $1'$\}, \{$10^6$, 0.75, $2'$\}, \{$5 \times 10^5$, 0.75, $2'$\}.

\begin{figure*}[htbp]
\includegraphics[width=\linewidth]{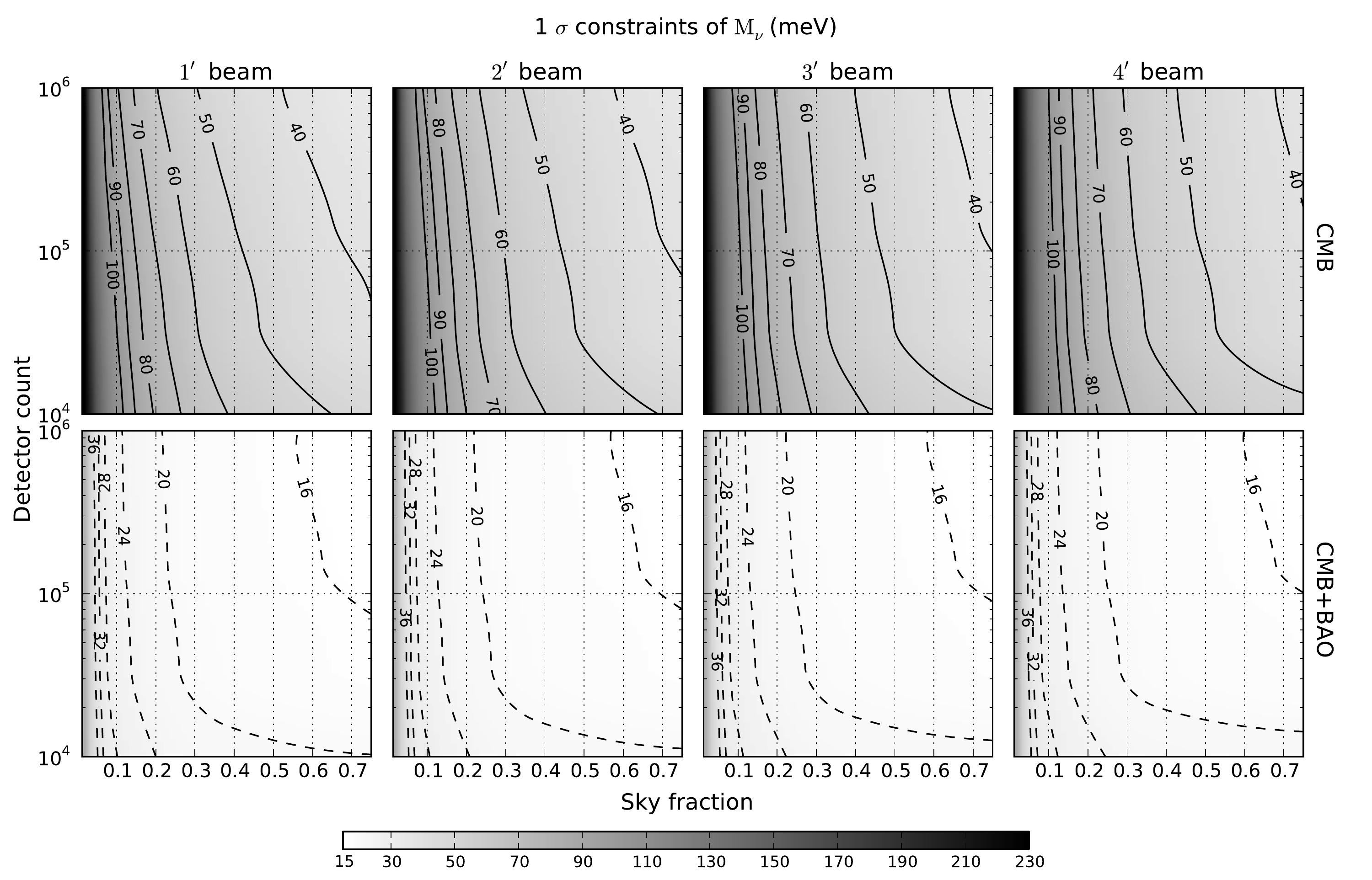}
\caption{1-$\sigma$ constraints on the total neutrino mass, for various detector numbers and observed sky fraction in units of meV. The top two panels show constraints from CMB for $1' - 4'$ beams. The bottom panels show constraints from CMB + BAO with the same beams in the CMB experiments. ``CMB'' includes lensing. We see that data from BAO pushes the constraints from CMB with lensing to a lower floor and a wide range of experimental configurations can obtain sub-20meV constraints on the sum of neutrino masses.}
\label{fig:sigmamnuXBAO_contour}
\end{figure*}

\textit{Relationship with Relic Helium Abundance $Y_P$ ---} In our analysis, we set the relic helium abundance $Y_P=0.248$, assuming that its value is measured externally. However, if we assume $N_{\rm eff}^{\rm BBN}$ is the same as $N_{\rm eff}^{\rm CMB}$, we can improve the constraints of $N_{\rm eff}$ at recombination by varying the value of $Y_P$ self-consistently for each $N_{\rm eff}^{\rm CMB}$ ($N_{\nu}$ in the expression below) and $\Omega_b h^2$ using the relation \cite{2008JCAP...06..016S}, 
\begin{equation}
Y_P = 0.2486 + 0.0016 [(\eta_{10} - 6)+ 100(S-1)]
\end{equation}
where $\eta_{10} = 273.9\, \Omega_B h^2$, $S = (1+7 \Delta N_{\nu}/43)^{1/2}$, and $\Delta N_{\nu} = N_{\nu} - 3$. 
With the imposed relation, we improve the constraints over the entire parameter space by 15-20\%. In particular, the best constraint for $N_{\rm eff}$ is 0.013, in the range of experimental configurations we consider in this work.

On the other hand, if $N_{\rm eff}^{\rm BBN}$ is different from $N_{\rm eff}^{\rm CMB}$ due to new physics, we can use CMB to constrain $N_{\rm eff}$ and $Y_P$ independently. In this case, the constraints on $N_{\rm eff}$ degrade from those listed in Table~\ref{table:Neff} by about a factor of 2, and the constraints on $Y_P$ is on the order of $10^{-3}$.

\subsubsection{$M_{\nu}$ forecast}
\label{ssec:m_nu}

Table~\ref{table:Mnu} presents a sample of $M_{\nu}$ constraints given different beam sizes, sky coverage, and detector counts in an experiment. The best constraint we get for CMB only is $M_\nu=34.1$ meV from the $10^6$ detector ($0.58\;\mu$K-arcmin equivalent), $f_{sky}=0.75$, $1'$ beam case. 
This constraint improves to $15.1$ meV when BAO is added. 
The top row of Fig.~\ref{fig:sigmamnuXBAO_contour} shows the trend of how neutrino constraints vary as a function of detector count, sky fraction, and beam sizes. 

\begin{table}[htbp]\centering
\ra{1.3}
\begin{tabular}{@{}lccccccccccl@{}}\hline
& \multicolumn{4}{c}{CMB} & \phantom{abc}&  \multicolumn{4}{c}{CMB+BAO}  \\
\cline{2-5} \cline{7-10} 
& $1'$&  $2'$  &  $3'$ & $4'$ &\phantom{abc}& $1'$ &  $2'$   &  $3'$  & $4'$   \\ \hline
$\mathbf{10^4}$ \textbf{detectors} \\
 $f_{sky}$ = 0.25  &  71.7 &  72.8 & 74.4 & 76.6 && 22.8 & 23.0 & 23.4 & 23.9 \\
 $f_{sky}$ =  0.50 & 54.7 & 55.7 & 57.2 & 59.2 && 20.6 & 20.9 & 21.3 &  21.9 \\
$f_{sky}$ =  0.75 &  48.1 & 49.0 & 50.5 & 52.5 && 20.1 &  20.4 & 20.9 &  21.5  \\
$\mathbf{10^5}$ \textbf{detectors} \\
 $f_{sky}$ = 0.25  & 63.3 & 65.2 & 66.7 & 68.3  && 19.7 & 19.8 & 19.9 & 20.1 \\
 $f_{sky}$ =  0.50 & 46.4 & 47.2 & 48.2 & 49.4  && 16.9 & 17.0 & 17.1 & 17.2 \\
$f_{sky}$ =  0.75 & 38.5 &  39.2 & 40.0 & 41.0 && 15.7 & 15.8 & 15.9 & 16.0 \\
$\mathbf{10^6}$ \textbf{detectors} \\ 
 $f_{sky}$ = 0.25 & 54.9 &  58.1 & 62.2 & 64.7 && 19.1 & 19.2 & 19.3 & 19.4   \\
$f_{sky}$ =  0.50 & 40.8 & 42.7 & 45.1 & 46.5 && 16.4 & 16.4 &  16.5 & 16.6 \\
 $f_{sky}$ = 0.75 & 34.1 & 35.7 & 37.2 & 38.3 && 15.1 & 15.2 &  15.3 & 15.3 \\
\hline
\end{tabular}
\caption{1-$\sigma$ constraints on $M_{\nu}$, in units of meV, from CMB and from CMB+BAO. ``CMB'' includes lensing. }
\label{table:Mnu}
\end{table}

For the levels of sensitivity achieved with $N_{\rm det} \geq 10^5$, constraints on $M_{\nu}$ are sample variance limited as opposed to sensitivity limited. We see from Fig.~\ref{fig:sigmamnuXBAO_contour} that as we increase detector counts, the constraints on $M_{\nu}$ reaches a plateau. For CMB alone, increasing $N_{\rm det}$ from $10^4$ to $10^5$ improves constraints for $f_{sky}$ = 0.5 cases by about 15\%, while increasing $N_{\rm det}$ from $10^5$ to $10^6$ improves constraints for the same cases by 6-12\% (smaller beam gives more improvement). 
In contrast, with CMB and BAO, increasing $N_{\rm det}$ from $10^4$ to $10^5$ improves constraints for $f_{sky}$ = 0.5 cases by about 20\%, while increasing $N_{\rm det}$ from $10^5$ to $10^6$ improves constraints by only 3\%. 
This shows that increasing $N_{\rm det}$ beyond $10^5$ does not improve much the constraints on $M_{\nu}$ once BAO is included. 
Decreasing beam sizes helps the CMB alone cases more when $N_{\rm det}$ is high, e.g. in $10^6$ $N_{\rm det}$ case, the improvement is almost 10\% for each decreased arc-minute. However this effect is washed away when BAO is added. For lower $N_{\rm det}$  regime, the improvement over decreasing beam size is a few \%.

\begin{figure}[!htbp]
\includegraphics[width=\linewidth]{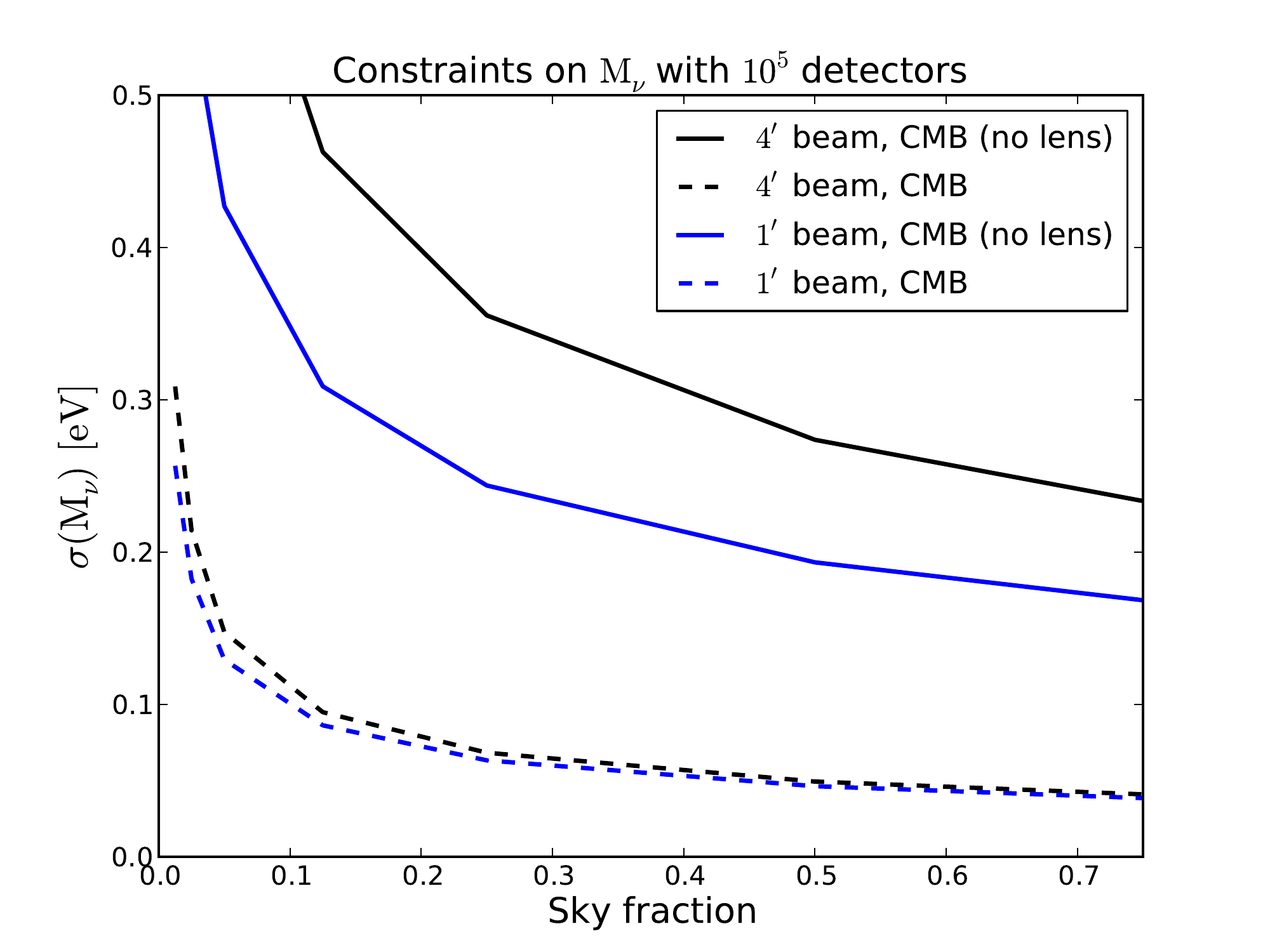}
\caption{Constraints on total neutrino mass across all sky fractions comparing CMB without lensing to CMB with lensing for experiments having $10^5$ detectors and $1'$ or  $4'$ beams. The constraints on $M_{\nu}$ are greatly improved by CMB lensing. This general trend is observed across all experimental configurations when the lensing power spectrum is added. }
\label{fig:simgamnu_lens_nolens}
\end{figure}

CMB lensing significantly improves the constraints on  $M_{\nu}$ from using primary spectra alone. Fig.~\ref{fig:simgamnu_lens_nolens} compares the constraints on  $M_{\nu}$ with and without using CMB lensing from an experiment with $10^5$ detectors and $4'$ beam. This is because the lensing spectrum breaks the degeneracy that neutrino mass has with $\Omega_c h^2$ in the primary spectra. 
Adding BAO information further breaks degeneracies as seen in the right column of  Table~\ref{table:Mnu}
The bottom row of Fig.~\ref{fig:sigmamnuXBAO_contour} shows that a wide range of experimental configurations can lead to tight neutrino mass constraints. Specifically, constraints go below $25$ meV for $f_{sky}\geq 0.125$, and for any number of detectors and beam sizes. 
These constraints are similar to those estimated by \citet{2013arXiv1308.4164F} for \textit{Planck} CMB plus DESI BAO plus redshift-space distortions and/or lensing measurements based on galaxies, e.g., DES/LSST.

\section{Dark-energy equation of state}
\label{sec:DarkEnergy}
\subsection{Overview}

The observed acceleration in the expansion of the universe is currently explained phenomenologically by dark energy. In this section, we focus on the constraints for a homogeneous dark energy with equation of state parametrized as 
\begin{eqnarray}
	\centering
		w(a) = w_0 + (1-a)\,w_a,
	\label{eq:wa_def}
\end{eqnarray}
allowing for a space and time varying dark energy field. In this parametrization, a cosmological constant would be the same as dark energy in its functional form when the equation of state $w(a)$ is constant and set to $-1$, i.e. $w_0 = -1$ and $w_a = 0$. Two questions ensues: (1) how well could future experiments differentiate a cosmological constant from inhomogenous dark energy? (2) how does dark energy evolve, if it does? This later question is the first step to capturing the much broader class of possible dark energy models. Once we precisely measure the equation of state, we will have a much better sense for directions in the theoretical modeling.

The effects of dark energy on cosmological observations are encapsulated in the Hubble parameter $H \equiv \dot{a}/{a}$, where $a = a(t)$ is the scale factor. The Friedmann equation relates the expansion rate of the universe with the energy densities of the contents in it:
\begin{equation}
\begin{split}
H(a)^2 =  H_0^2  (& \Omega_r a^{-4} + \Omega_M a^{-3} +  \Omega_K a^{-2} +\\
                                 &  \Omega_{DE} \exp \left(3 \int_a^1 \frac{da' }{a'} [1+w(a')] \right) ),
\end{split}
\end{equation}
where $\Omega_r$, $\Omega_{M} $, $\Omega_{DE} $, and $\Omega_{K} $ are the density of radiation, matter, dark energy, and curvature respectively, and $w(z)$ is the specific parametrization of the equation of state for dark energy. 

The Hubble expansion rate $H$ also governs the growth of linear perturbations, $\delta$, in structure formation:
\begin{equation}
\ddot{\delta} + 2H \dot{\delta} - 4 \pi G \rho_M \delta = 0.
\end{equation}

Primordial CMB fluctuations constrain dark energy parameters through the effects $H$ have on them at recombination, i.e. $z \simeq 1100$ -- acoustic peaks positions are sensitive to the value of $w$ \cite{2001PhRvD..64l3527H}. If $w < -1$,  both $C_\ell^{TT}$ and $C_\ell^{EE}$ would shift towards higher $\ell$ and if $w > -1$, they would shift towards lower $\ell$. The $w_a$ parameter gives the same effects as $w_0$ but at a smaller amplitude. 
One gets the same qualitative feature when $H_0$ (value of $H(z)$ at $z$=0) is decreased or increased respectively. Therefore, when we only use information from the CMB ($z \simeq 1100$), dark-energy equation of state parameters are degenerate with geometrical parameters, like $H_0$ and $\Omega_K$. 

\begin{figure*}[htbp]
\includegraphics[width=\linewidth]{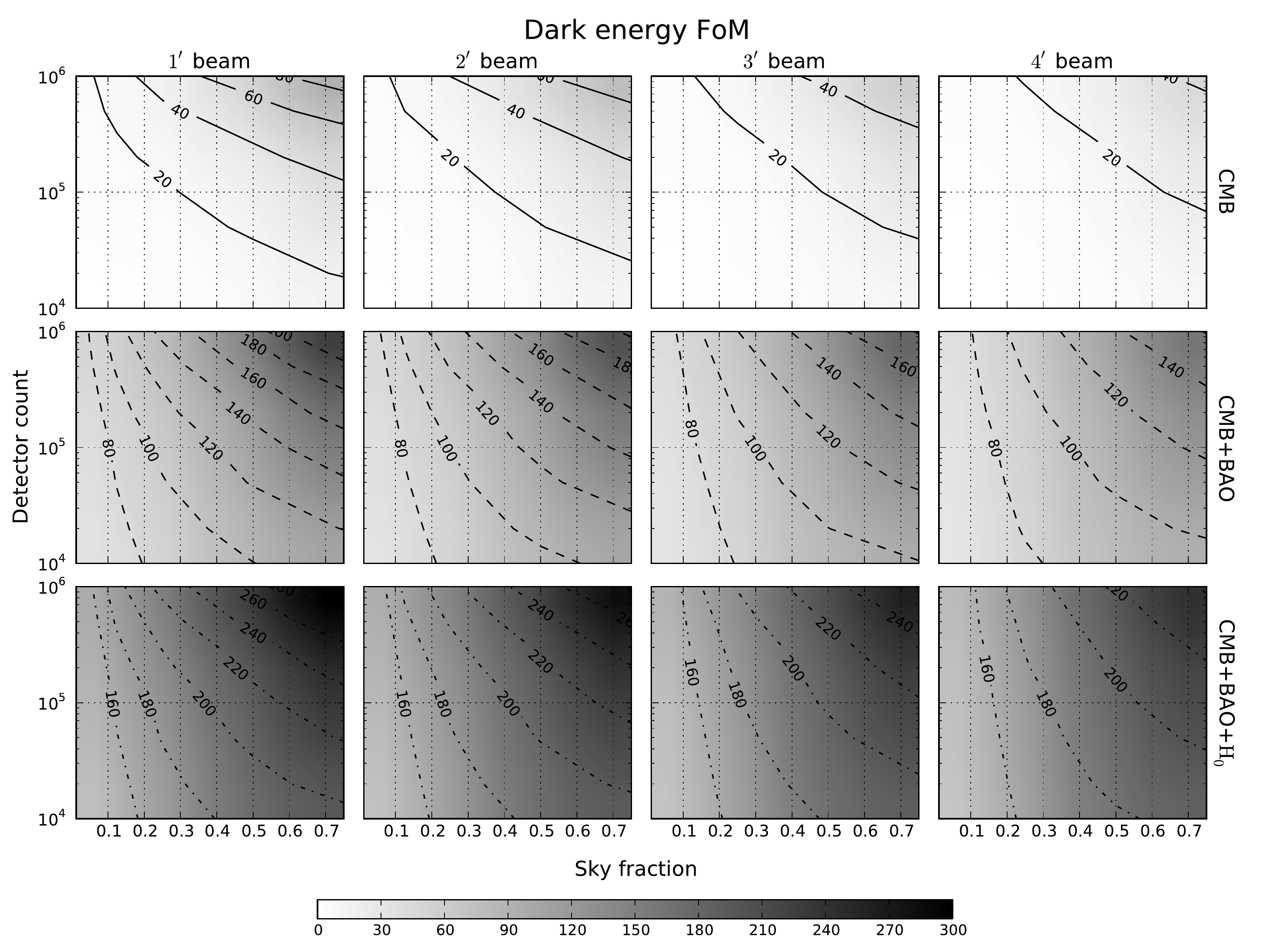}
\caption{The Dark Energy Task Force Figure of Merit (DETF) (neutrino mass included in Fisher space) as a function of detector number counts and sky fraction for $1' - 4'$ beam size. The top row shows FoM from CMB. The middle row shows FoM from CMB+BAO. The bottom row shows FoM from CMB+BAO + $1\%$ $H_0$ prior. ``CMB'' includes lensing. The obvious trend is that the more detectors and sky coverage, and the smaller the beam, the better the DETF FoM is. However, the case of a CMB+BAO+$H_0$ prior combination shows that different beams have very close performances.}

\label{fig:DEFOM_contour}
\end{figure*}

To break this degeneracy, low redshift probes like weak lensing (CMB and optical), BAO, and $H_0$ measurements are essential. Here we focus on CMB lensing in relation to dark energy. CMB lensing is sensitive to structures at various redshifts (the lenses) and the CMB (the source), cf. Ref.~\cite{2010GReGr..42.2197H}. Because they are both a function of $H(z)$, we can observe the effects of dark energy through them. The effect of the equation of state $w$ on the lensing power spectrum is an overall enhancement or suppression in power across all scales with a minor scale dependence. Because CMB lensing probes redshifts higher than optical surveys, it is particularly useful for studying structure formation beyond $z > 2$ -- a direct observatory of the early history of dark energy. Specifically, if effects of a non-standard dark energy manifest at high $z$ ({\it e.g.} $z \sim 5$), then structure formation could be significantly suppressed, which CMB lensing can uniquely constrain.

\subsection{Results and discussion}

As discussed in the previous section, to constrain the dark-energy equation of state, it is essential to have observations from various redshifts and multiple probes. In this section, we present the constraints on $w_0$, $w_a$, and the Figure of Merit (FoM) as defined by the Dark Energy Task Force (DETF) \cite{2006astro.ph..9591A}, from the CMB, with and without $H_0$ prior, and with and without BAO measurements from DESI.

In addition to constraints on $w_0$ and $w_a$, and the DETF-FoM, we quote constraint on $w_p$ at $a_p$ where $a_p$ is the value of the scale factor $a$ when the uncertainties of $w(a)$ are minimized \cite{2006astro.ph..9591A}. To be explicit, FoM = $1/(\sigma(w_a) \sigma(w_p))$. 
Table \ref{table:DEparams} in the Appendix lists the constraints from CMB, CMB+BAO, and CMB+BAO + $1\%$ $H_0$ prior. CMB includes CMB lensing. 
The best FoM is 303. As defined earlier, this Fisher matrix parameter space includes massive neutrinos. For reference to the original DETF FoM, we include the constraints with fixed sum of neutrino mass in table \ref{table:DEparams_fixedMnu}, in the Appendix. In this scenario, the best FoM is 576.

In contrast to the rest of the models investigated in this work, the DETF-FoM is sensitivity limited. 
For example, for CMB+BAO+$H_0$ prior $1'$ beam cases, increasing $N_{\rm det}$ from $10^4$ to $10^5$ improves the FoM by 10-19\%, while increasing $N_{\rm det}$ from $10^5$ to $10^6$ improves the FoM by 21-29\% (bigger $f_{sky}$, more improvement). 
For CMB alone and $1'$ beam, increasing $N_{\rm det}$ from $10^4$ to $10^5$ and  from $10^5$ to $10^6$  improves the FoM by more than double respectively. 
Improvements from decreasing beam sizes become more noticeable at high $N_{\rm det}$, and the relative improvement per arc-minute decreases as beam size decreases. The improvement ranges from a few \% to tens of \%.
While $f_{sky}$ does not drive the improvement of FoM as much, the FoM does improve for increasing $f_{sky}$ for all cases in the table, though the sensitivity of the experiment decreases.

\begin{figure}[htbp]
\includegraphics[width=\linewidth]{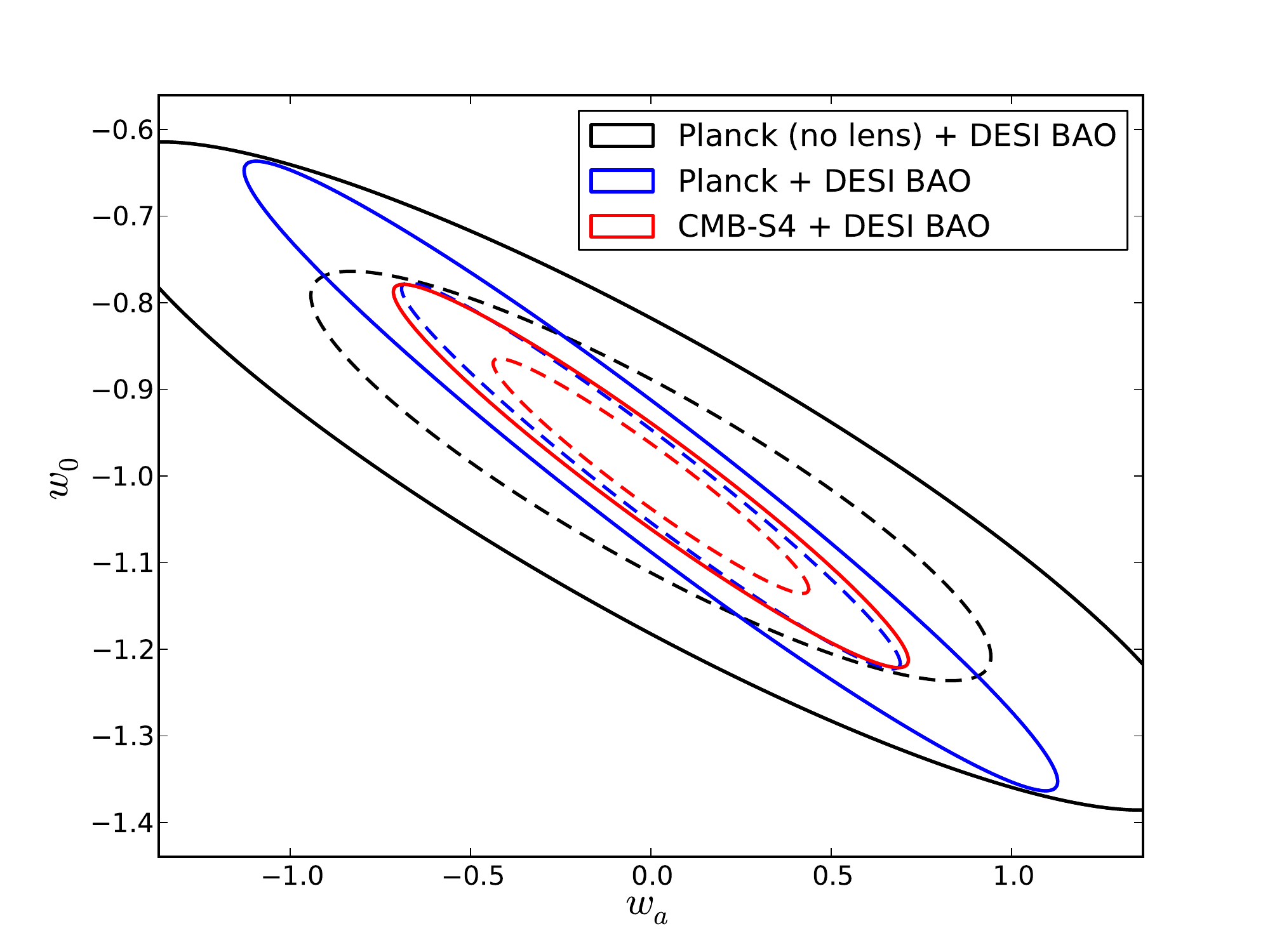}
\caption{The error ellipses for $w_0$ and $w_a$ comparing CMB-S4 and \textit{Planck} given BAO information from DESI and marginalizing over $M_{\nu}$. The FoM from \textit{Planck} (no lensing) + DESI BAO is 22, and that from CMB-S4 + DESI BAO is 141. This CMB-S4 experiment has 500,000 detectors, $3'$ beam, and covers half of the sky. }
\label{fig:ellipse_w0_wa}
\end{figure}

Fig.~\ref{fig:DEFOM_contour} shows the trend of how the FoM improves with increasing detector counts and sky fraction. We note that BAO and $H_0$ prior improve the FoM from using CMB alone from a factor of several to an order of magnitude. Also, with BAO and $H_0$ prior, increasing the detector number does not improve the FoM much at $f_{sky}$ smaller than 0.2.

Fig.~\ref{fig:ellipse_w0_wa} illustrates how much a CMB-S4 experiment improves on the constraints on $w_0$ and $w_a$ over \textit{Planck} including DESI BAO and marginalizing over neutrino masses. The FoM from \textit{Planck} (no lensing) + DESI BAO is 22, and that from CMB-S4 + DESI BAO is 141. In this figure, the assumed CMB-S4 configuration has 500,000 detectors, $3'$ beam, and covers half of the sky. The improvement comes mainly from the tight constraints one gets for neutrino masses from CMB-S4 and DESI BAO. 

To further constraint dark-energy parameters, adding extra probes like supernovae and optical weak lensing, as suggested by the DETF report, will certainly help.


\section{Dark-matter annihilation}
\label{sec:DarkMatter}
\subsection{Overview}

If Dark Matter (DM) is a weakly interacting massive particle (WIMP) and it is a thermal relic, then its self-annihilation cross-section can be determined by its relic density today through the Boltzmann equation \cite{1986PhRvD..33.1585S, 1991NuPhB.360..145G, 2012PhRvD..86b3506S}. 

Depending on the model, DM particles can annihilate into gauge bosons, charged leptons, neutrinos, hadrons, or more exotic states. These annihilation products then decay or interact with the photon-baryon fluid and produce electrons, positrons, protons, photons, and neutrinos. Neutrinos do not further interact with the photon-baryon fluid, but interact gravitationally and their effects can be observed through the lensing of the CMB. A proton's energy deposition to the fluid is inefficient due to their high penetration length. Hence, the main channels for energy injection are through electrons, positrons, and photons. At high energies, positrons lose energy through the same mechanisms as electrons. High energy electrons lose energy through inverse Compton scattering of the CMB photons, while low energy electrons lose energy through collisional heating, excitation, and ionization. Photons lose energy through photoionization, Compton scattering, pair-production off nuclei and atoms, photon-photon scattering, and pair-production through CMB photons \cite{1989ApJ...344..551Z}. The rate of energy release per unit volume by a self-annihilating DM particle is given by \cite{2009PhRvD..80b3505G}
\begin{equation}
\frac{dE}{dt}(z) = \rho^2_c c^2 \Omega^2_{DM} (1+z)^6 f \frac{\langle \sigma v \rangle}{m_{DM}},
\end{equation}
where $\rho_c$ is the critical density of the universe today, $\Omega_{DM}$ is the DM density, $f$ is the energy deposition efficiency factor, $\langle \sigma v \rangle$ is the velocity-weighted annihilation cross section, and $m_{DM}$ is the mass of the DM particle, assumed to be a Majorana particle in this work.

Due to these processes, the photon-baryon plasma is heated and the ionization fraction is modified. This leads to modifications of the recombination history \cite{2005PhRvD..72b3508P, 2009PhRvD..80b3505G} and, consequently, of the CMB spectra. For details on the energy injection processes, see~\citet{2009PhRvD..80d3526S}.  The energy injection due to DM annihilation broadens the surface of last scattering, but does not slow recombination \cite{2005PhRvD..72b3508P}.
The extra scattering of photons at redshift $z \lesssim 1000$ damps power in the CMB temperature and polarization fluctuations at small angular scales ($\ell \gtrsim100$),  and adds power in the E-mode polarization signal at large scales ($\ell \lesssim 100$). The ``screening" effect at $\ell \gtrsim 100$ goes as an exponential suppression factor $C_\ell\rightarrow e^{-2\Delta\tau} C_\ell$, where $\Delta\tau$ is the excess optical depth due to dark-matter annihilation. This exponential factor is partially degenerate with the amplitude of the scalar perturbation power spectrum $A_s$, and polarization data helps to break this partial degeneracy.

\begin{figure}[htbp]
\includegraphics[width=\linewidth]{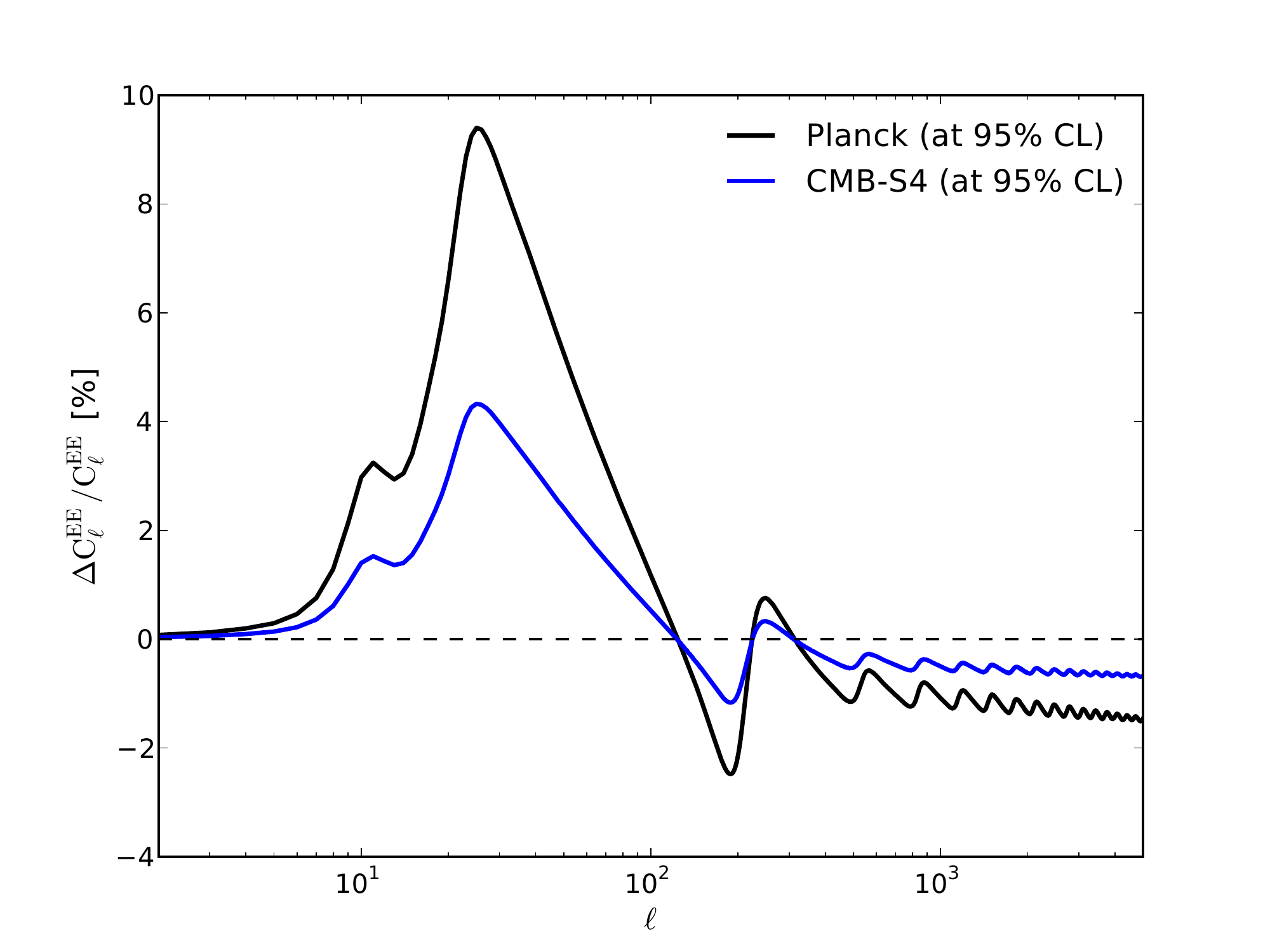}
\caption{Percentage relative deviation of $C_\ell^{EE}$ from fiducial cosmology when DM annihilation is taken into account. The \textit{Planck} and CMB-S4 lines illustrate the power of each experiment to differentiate a model with DM annihilation from the fiducial cosmology at the $95\%$ C.L..}
\label{fig:CEE_DM_Plk_cmbS4}
\end{figure}

The CMB is sensitive to the parameter $p_{ann}$, defined as
\begin{equation}
p_{ann} \equiv f \frac{\langle \sigma v \rangle}{m_{DM}}.
\end{equation}
In this work, we take a thermal cross-section $\langle \sigma v \rangle = 3 \times 10^{-26} \rm  cm^3 .s^{-1}$ and $f = 1$, and we will present the 95\% C.L. upper limit of the DM particle mass, $m_{DM}$, that a CMB Stage-IV experiment could reach. We note that $\langle \sigma v \rangle$ could vary for different ranges of $m_{DM}$ \cite{2012PhRvD..86b3506S} and that the value of $f$ is a function of redshift and interaction of annihilation products~\cite{2013PhRvD..87l3513S}. The reader can easily scale our results for any other models considered.

In particular, we expect most of the constraining power of CMB-S4 to come from its polarization spectrum $C_\ell^{EE}$ and the cross spectrum $C_\ell^{TE}$. Fig.~\ref{fig:CEE_DM_Plk_cmbS4} illustrates the percentage deviation from the fiducial cosmology in $C_\ell^{EE}$ at the 95\% C.L. value of $p_{ann}$ that \textit{Planck} and CMB-S4 would be able to differentiate respectively. The CMB-S4 configuration chosen for this figure has $10^6$ detectors, $f_{sky} = 0.75$, and $1'$ beam.

\subsection{Results and discussion}
In the range of sensitivities and beam sizes considered, we found that the main factor that improves the limit in $m_{DM}$ is sky coverage $f_{sky}$. This means that the constraints are largely sample variance limited. 

\begin{figure}[htbp]
\includegraphics[width=\linewidth]{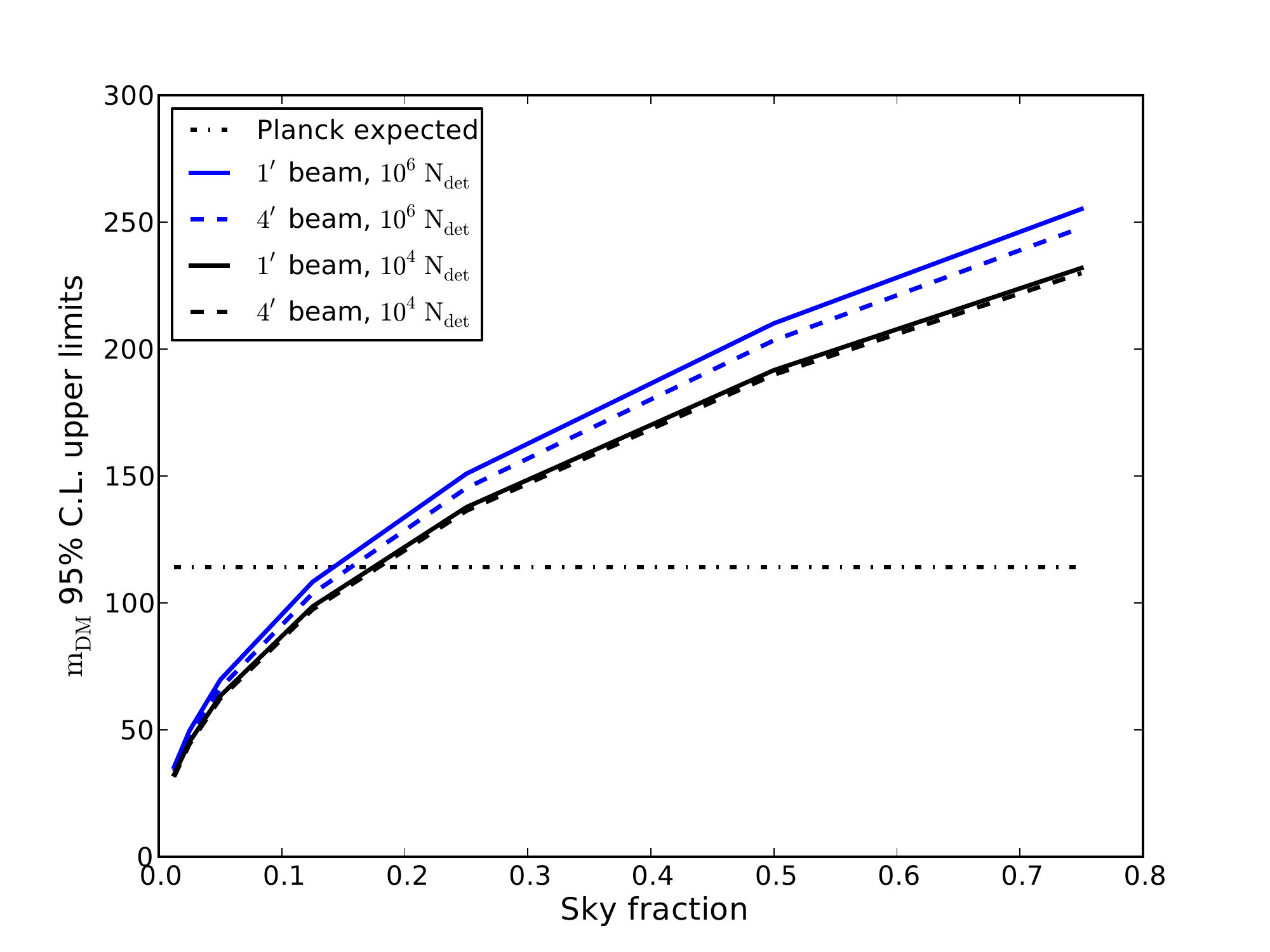}
\caption{95\% C.L. upper limit for $m_{DM}$ in GeV as a function of $f_{sky}$ for a few experimental configurations. The blue(black) lines correspond to $10^6$ ($10^4$) detectors and the solid(dash) lines correspond to $1'$($4'$) beams. The dashed-dotted line shows the limit expected from \textit{Planck}. While it matters more to have a small beam when there are a lot of detectors, the dependence is small compared to the dependence on sky fraction. Therefore, to have a tighter constraint on $m_{DM}$, we need as much sky coverage as possible. }
\label{fig:mDM_vs_fsky}
\end{figure}
Fig.~\ref{fig:mDM_vs_fsky} illustrates how steep the dependence is on $f_{sky}$ (it goes as $\sim f_{sky}^{1/2}$), and the mild dependence on detector number and beam size.

It remains exciting to see significant improvement on $m_{DM}$ limits from \textit{Planck} given a range of possible configurations of CMB-S4. At perfect energy deposition $f = 1$, \textit{Planck} could exclude $m_{DM} <\,114$ GeV at $95\%$ C.L. for $\langle \sigma v \rangle = 3 \times 10^{-26} \rm cm^3 s^{-1}$. Any CMB-S4 configuration with beams smaller than $4$', minimum of $10^4$ detector, covering half the sky could exclude $m_{DM} <\, 190$ GeV at the $95\%$ C.L. or better. In particular, the best exclusion limit considered here is $255$ GeV with $10^6$ detectors, $1'$ beam size, and $f_{sky}$=0.75. Configurations that give $95\%$ C.L. upper limits higher than $\sim 200$ GeV are sensitive to $100$ GeV at the $3\sigma$ level, where many of the current direct detection experiments tune their sensitivities.

\section{Inflation}
\label{sec:Inflation}
\subsection{Overview}

The theory of cosmic inflation was proposed to solve the missing monopole problem, the flatness problem, and the horizon problem from the standard Big Bang theory \cite{1981PhRvD..23..347G}. Inflation invokes a period of rapid expansion of the universe before the standard Big Bang expansion phase. Because of the rapid expansion, any relics would be diluted to a point where they would be extremely rare in our observable universe. Similarly, the curvature of the Universe could be diluted as well. For the horizon problem, modes that would not have been in causal contact in standard Big Bang were in fact once in causal contact in the inflationary framework before their horizon exit \cite{dodelson:2003}. Outside the horizon, their amplitudes were frozen. Therefore, after they re-entered the horizon, modes that would not have been in causal contact if not for inflation appear to have equilibrated with each other. Not only does inflation solve these problems, it sets the initial conditions for large-scale structures, thus providing a physical foundation for the observed fluctuations in the CMB and LSS.
 
During inflation, quantum fluctuations were stretched and became classical perturbations. Scalar perturbation of the metric seeded the formation of large scale structure. Tensor perturbation of the metric created primordial gravitational waves. The observables caused by these perturbations provide definite signatures to confirm or falsify the theory of inflation. To study them, we parameterize the perturbation spectra using a power law in wavenumber $k$, as follows:

The scalar spectrum is
\begin{equation}
P_s(k) = A_s(k_*)\left(\frac{k}{k_*}\right)^{n_s(k_*)-1+\frac{1}{2}\alpha_s(k_*)\log(k/k_*)},
\end{equation}
where $A_s$ is the amplitude of the scalar spectrum, $n_s$ is the scalar spectral index, $k_{*}$ is the pivot scale, and $\alpha_s$ is the running of $n_s$. 
The tensor spectrum can be written as:
\begin{equation}
P_t(k) = A_t(k_*)\left(\frac{k}{k_*}\right)^{n_t(k_*)},
\end{equation}
where  $A_t$ is the amplitude of the tensor spectrum and $n_t$ is the tensor spectral index.

We can measure $A_s$, $n_s$, $\alpha_s$, $A_t$, and $n_t$ using the CMB. 
One of the most sought-after parameter, the tensor-to-scalar ratio $r\equiv P_t/P_s$, tells us the energy scale of inflation. Specifically, 
\begin{equation}
V^{1/4} = 1.06 \times 10^{16} {\rm GeV} \left( \frac{r}{0.01} \right)^{1/4}
\end{equation}
Therefore, a value of $r$ larger than $0.01$ would confirm that inflation happened at an energy scale comparable to that at the Grand Unified Theory, at which the strong, weak and electromagnetic forces are unified. The CMB provides a unique window to the high energy physics that cannot be attained in terrestrial accelerator experiments. 

The value of $r$ is also directly related to how big the field excursion $\Delta \phi$ is from when fluctuations seen in the CMB where created to the end of inflation. Given the number of $e$-folds $N_e$, in single-field slow-roll and single-field DBI inflation \cite{2004PhRvD..70j3505S,1997PhRvL..78.1861L}, we can write \cite{1997PhRvL..78.1861L}

\begin{equation}
N_e = \int \frac{da}{a} = \int H dt = \int \frac{HM_{pl}}{\dot{\phi}}\frac{d\phi}{M_{pl}} = \sqrt{8} r^{-1/2} \frac{\Delta \phi}{M_{pl}},
\end{equation}
where $M_{pl}$ is the Planck mass, $r = P_t/P_s$, $P_t \propto H^2/M_{pl}^2 $, and $P_s \propto H^4/\dot{\phi}^2$. For $N_e \sim 30$, the lower bound that corresponds to a minimal reheating temperature \cite{2005JCAP...05..008E}, a Planck mass field range corresponds to $r \sim 0.01$. (Similarly, for $N_e \sim 60$, $\Delta \phi = M_{pl}$ implies $r \sim 0.002$).  Super-Planckian field excursion, $\Delta \phi > M_{pl}$ implies that the inflaton field is sensitive to an infinite series of operators of arbitrary dimensions \cite{2009AIPC.1141...10B}.  In order for a large-field inflation model to be UV complete, effective field theory requires a shift symmetry in the field, which protects the flatness of the potential over a large field range. Differentiating between sub-Planckian and super-Planckian field excursion during inflation would rule out different classes of inflationary models. 

Here we use the simplest class of models -- single-field slow-roll inflation to illustrate how observations can falsify inflationary models. The conditions for slow-roll require the inflationary potential to be flat enough such that the slow-roll parameters $\epsilon$ and $\eta$ are small. They are defined as follows: 

\begin{eqnarray}
\epsilon  &\equiv& \frac{M^2_{pl}}{2}\left(\frac{V_{,\phi}}{V}\right)^2 \ll 1 \\
\eta &\equiv& M^2_{pl} \frac{V_{,\phi\phi}}{V} \ll 1
\end{eqnarray}
where $V$ is the inflaton potential, and the subscript $,\phi$ denotes the partial derivative with respect to $\phi$. 

We can write $n_s$, $\alpha_s$, $n_t$, and $r$, to leading order, in terms of the slow-roll parameters:
\begin{eqnarray}
n_s &=&1- 6\epsilon +2\eta  \\
n_t &=& -2\epsilon \\
\alpha_s &= & -16\epsilon\eta + 24\epsilon^2 + 2\xi^2 \\
r &=& 16\epsilon,
\end{eqnarray}
where $\xi^2 = M^4_{pl} V_{,\phi} V_{,\phi\phi\phi}/V^2$.

Therefore, for each specific single field slow-roll model, there exists a unique set of predictions on each of these parameters. This allows us to rule out models by measuring the values of the parameters.

Inflation predicts an almost but not exactly flat universe, where $|\Omega_K|$ is on the order of $10^{-4}$ \cite{2013arXiv1303.5082P, 2013arXiv1309.5381A} . This is a consequence of the large scale modes at the horizon, which contribute to spatial curvature. A statistically significant deviation from the expectation of inflation for the curvature would give us information on the process of inflation. 
For example, if significant departure from flatness is measured, it can mean that inflation was not slow-rolling when perturbations of the scales just larger than our observable universe exited the inflationary horizon. 

The current best constraints for these parameters are from the first release of \textit{Planck} data \cite{2013arXiv1303.5082P}. It is measured that $n_s = 0.9603 \pm 0.0073$ (and $n_s = 0.9629 \pm 0.0057$ when combined with BAO), $\alpha_s =  -0.0134 \pm 0.0090$, $\Omega_K = -0.058^{+0.046}_{-0.026}$ with \textit{Planck} and \textit{WMAP} polarization data (and $\Omega_K = -0.0004 \pm 0.00036$ when combined with BAO). An upper bound for r set at $r < 0.11$ at 95\% C.L.

To get to the next qualitatively significant level of constraints for these parameters or confirm null results, we need to at least constrain $r$ to 0.002 at 1-$\sigma$. This will confirm a discovery for large-field inflation in the case where $r \gtrsim 0.01$ if $N_e \sim 30$. Alternatively, a null result at this level will rule out large field inflation. For curvature $\Omega_K$, it would be very interesting to get 1-$\sigma$ constraints at the level of $10^{-5}$. While $n_s$ is constrained to a level where we confirm inflation predictions, more precise constraints will be needed to rule out models. As for $\alpha_s$, most inflationary models predict it to be undetectable, so any detection would be interesting.

\subsection{Results: : $\Omega_K$, $n_s$, $\alpha_s$}
In this section, we present the 1-$\sigma$ constraints of $\Omega_K$, $n_s$, and $\alpha_s$ that are close to getting to evidence (5$\sigma$) and/or discovery (3$\sigma$) regimes for the grid of experimental setups. The treatment for constraining the tensor-to-scalar ratio $r$ is different from that for the other three parameters so we devote a separate section for $r$. 

\subsubsection{$\Omega_K$ forecast}

Table \ref{table:OmegaKconstraints} lists the constraints for $\Omega_K$ using CMB combined with BAO measurements and $1\%$ $H_0$ prior. CMB experimental inputs listed have $10^4$, $10^5$, and $10^6$ detectors, $1' - 4'$ beams, and fixed sky fraction of 0.75. 
\begin{table}[h]\centering
\ra{1.3}
\begin{tabular}{@{}lp{0.25cm}p{1cm} p{1cm}p{1cm}p{1cm}p{1cm}@{}}\hline
&&  $1'$ & $2'$ & $3'$ & $4'$\\ \hline
$\mathbf{10^4}$ \textbf{detectors} \\
CMB (no lens)  && 8.03 & 8.42  & 9.02  & 9.77  \\
CMB                   && 5.29 & 5.41  &  5.55 & 5.70 \\
CMB+$H_0$    && 1.30 & 1.31  & 1.33 & 1.36 \\
CMB+BAO        &&  0.94 &  0.96 &  0.98 &  1.01  \\
CMB+BAO+$H_0$ && 0.93  & 0.95  & 0.97  & 1.00  \\
$\mathbf{10^5}$ \textbf{detectors} \\
CMB (no lens)  && 6.26 & 6.61 & 7.09 & 7.71  \\
CMB                   &&   4.24 & 4.31  & 4.39 & 4.52 \\
CMB+$H_0$    && 1.21 & 1.23 & 1.24 & 1.26 \\
CMB+BAO        &&  0.84 & 0.86 & 0.88 &  0.90 \\
CMB+BAO+$H_0$ && 0.84 & 0.85 & 0.87 & 0.89 \\
$\mathbf{10^6}$ \textbf{detectors} \\
CMB (no lens)  && 4.89  & 5.38  & 5.96 &6.47  \\
CMB                    && 3.37 & 3.65 & 3.84 & 4.03 \\
CMB+$H_0$     && 1.13 & 1.16 & 1.19 & 1.22 \\
CMB+BAO         && 0.75  & 0.79 & 0.83 & 0.85 \\
CMB+BAO+$H_0$ && 0.74 & 0.78 & 0.82 & 0.84 \\
\hline
\end{tabular}
\caption{$\Omega_K$ 1-$\sigma$ constraints in units of $10^{-3}$ for 5 combinations of CMB with DESI BAO and 1\% $H_0$ prior at $f_{sky}$ = 0.75.}
\label{table:OmegaKconstraints}
\end{table}

Constraints on $\Omega_K$ are sample variance limited in the sensitivity range we consider for CMB-S4 in this study. Increasing $N_{\rm det}$ by an order of magnitude in the CMB only improves $\Omega_K$ constraints by about 20\%, while adding BAO and/or $H_0$ improves the constraints by factor of about 5. Decreasing beam sizes improves the constraints at percent level. 

The effects of the curvature density $\Omega_K$ on the CMB temperature and E-mode polarization power spectra are degenerate with those from $H_0$ \cite{1999MNRAS.304...75E, 2006ApJ...650L..13H}, because a larger correlation angle between hot spots could be due to both curvature of space and the surface of last scattering being closer to us. Measuring $\Omega_K$ at multiple redshift slices could break this degeneracy. Therefore, we will use multiple probes -- CMB lensing, BAO, $H_0$ priors -- for constraining $\Omega_K$ in this section. 

\begin{figure}[htbp]
\includegraphics[width=\linewidth]{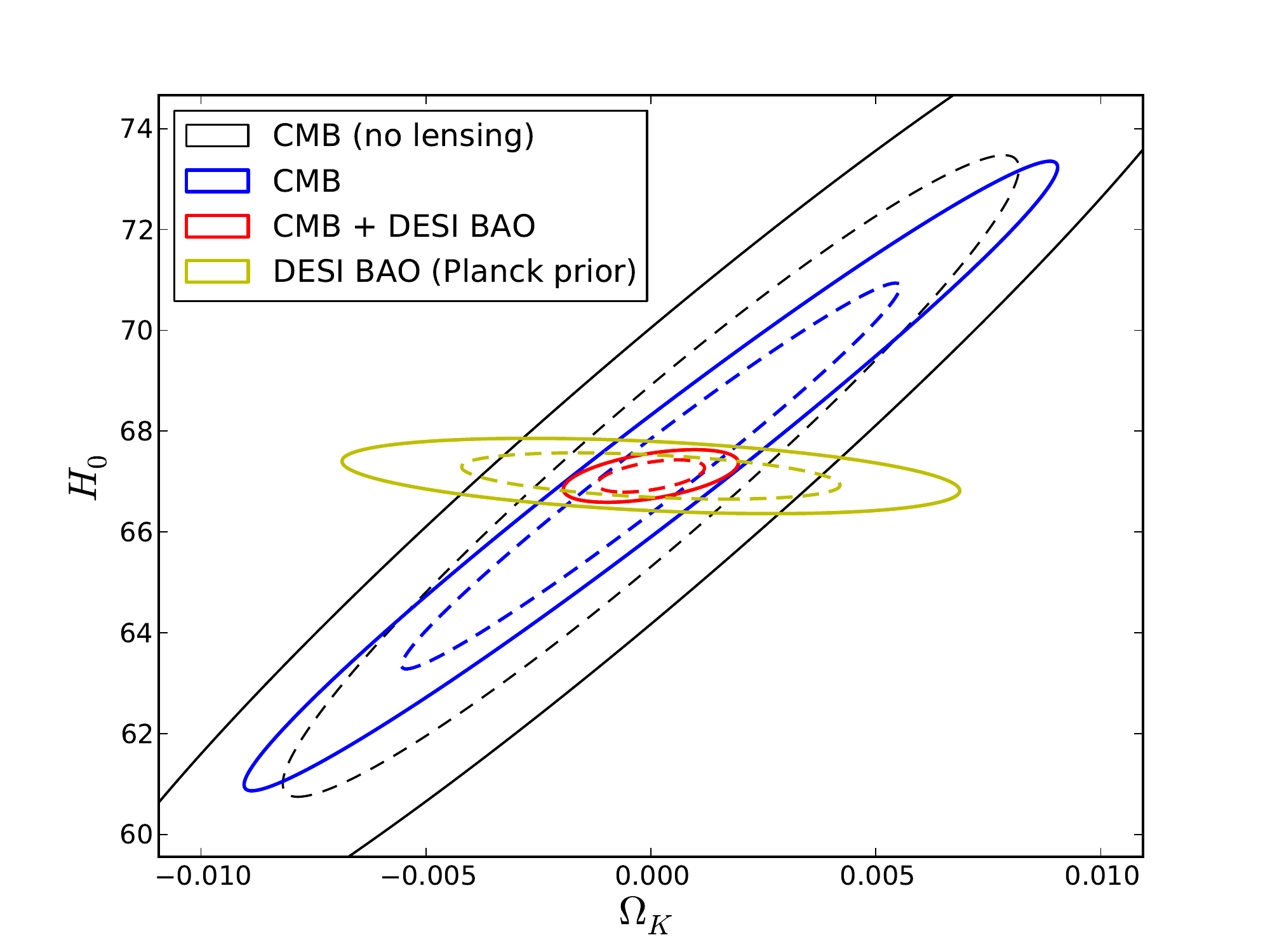}
\caption{The 1-$\sigma$ (dash line) and 2-$\sigma$ (solid line) constraint ellipses of $\Omega_K$ vs. $H_0$. This plot shows how having the BAO handle breaks degeneracies between the two parameters.}
\label{fig:OmKhubbleEllipse}
\end{figure}

\begin{figure}[htbp]
\includegraphics[width=\linewidth]{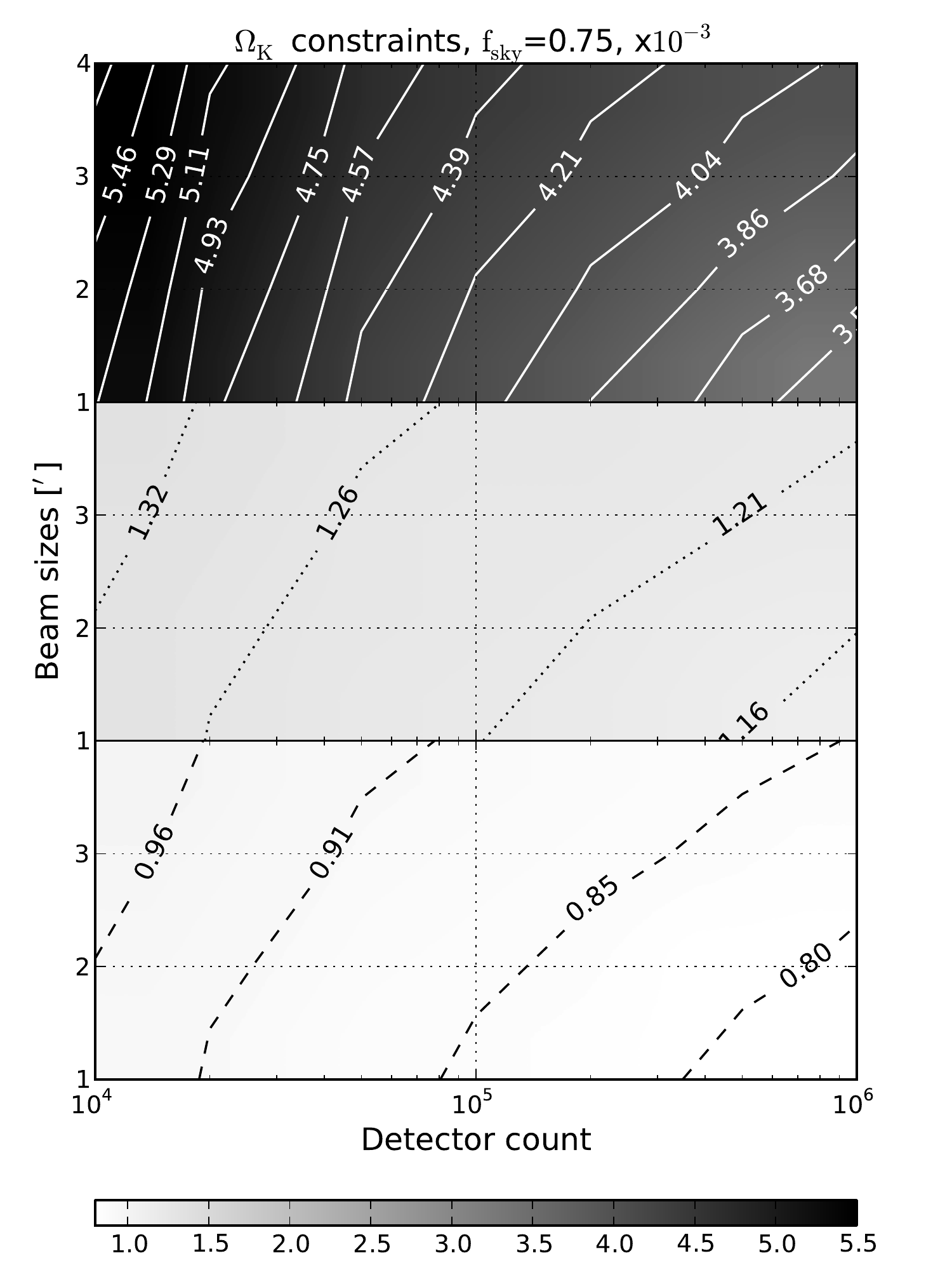}
\caption{The 1-$\sigma$ constraints of $\Omega_K$$\times 10^{-3}$ in the plane of beam sizes and detector counts for fixed $f_{sky}$ = 0.75. We show constraints from CMB, CMB+1\%$H_0$ prior, and CMB+BAO from top to bottom panel. ``CMB'' includes lensing. BAO provides about an order of magnitude improvement in the whole space by breaking degeneracies, better than adding a 1\% $H_0$ prior to CMB data.}
\label{fig:OmK_contour_all}
\end{figure}

CMB lensing provides an overall factor of $\sim$ 2 improvement from CMB (TT, EE, TE) for the smaller beam cases, and about a factor of three improvement for $8'$ beam. When BAO information is added to CMB lensing, there is a factor of 5 to 10 improvement across the whole grid. This is because the BAO signal is orthogonal to the CMB in $\Omega_K$ and $H_0$ space, as illustrated in Fig.~\ref{fig:OmKhubbleEllipse}. We expect the Hubble parameter to be constrained to about or better than 1\% in the coming decades. The $\Omega_K$ constraints improve by about a factor of $3$ to $5$ from CMB lensing given the 1\% $H_0$ prior. Fig.~\ref{fig:OmK_contour_all} shows the 
trends of the constraints in $\Omega_K$ for CMB, CMB+$1\%$ $H_0$ prior, and CMB+BAO respectively (CMB includes lensing).

From Table~ \ref{table:OmegaKconstraints}, we note that the 1-$\sigma$ constraints on $\Omega_K$ is 0.00075 from CMB+DESI BAO, which is very close to the regime where we want to be able to constraint $\Omega_K$. To further shed light on the mean spatial curvature of our universe, 21-cm mapping \cite{2008PhRvD..78b3529M}, QSOs, and the Lyman-$\alpha$ forest will also be useful \cite{2013arXiv1308.4164F}. 

\subsubsection{$n_s$, $\alpha_s$ forecast}

Table~\ref{table:n_s} summarizes the constraints for $n_s$, the scalar spectral index, for experiments with $10^4$, $10^5$, and $10^6$ detectors, $1'-4'$ beams, covering $25-75\%$ of the sky, lensing included. We can also see that the constraints are similar across beam sizes and number of detectors.

\begin{table}[htbp]\centering
\ra{1.3}
\begin{tabular}{@{}lccccccccccl@{}}\hline
& \multicolumn{4}{c}{CMB} & \phantom{abc}&  \multicolumn{4}{c}{CMB+BAO}  \\
\cline{2-5} \cline{7-10} 
& $1'$&  $2'$  &  $3'$ & $4'$ &\phantom{abc}& $1'$ &  $2'$   &  $3'$  & $4'$   \\ \hline
$\mathbf{10^4}$ \textbf{detectors} \\
 $f_{sky}$ = 0.25 & 2.91 & 2.94 & 2.98 & 3.04 && 2.19 & 2.23 & 2.29 & 2.36  \\
 $f_{sky}$ = 0.50 & 2.11 & 2.13 & 2.16 & 2.21 && 1.64 & 1.67 & 1.71 & 1.75  \\
 $f_{sky}$ = 0.75 & 1.76 & 1.77 & 1.80 & 1.83 && 1.39 & 1.42 & 1.45 & 1.48  \\
$\mathbf{10^5}$ \textbf{detectors} \\
 $f_{sky}$ = 0.25  & 2.66 & 2.73 & 2.80 & 2.86  && 1.93 & 1.98 & 2.04 & 2.12  \\
 $f_{sky}$ =  0.50 & 1.94 & 1.97 & 2.01 & 2.06  && 1.44 & 1.47 & 1.51 &1.56 \\
$f_{sky}$ =  0.75 & 1.60 & 1.63 & 1.66 & 1.70 && 1.22 & 1.24 & 1.28 & 1.32 \\
$\mathbf{10^6}$ \textbf{detectors} \\ 
 $f_{sky}$ = 0.25 & 2.38 &  2.48 & 2.62 & 2.73 && 1.70 &  1.76 & 1.83 & 1.92   \\
$f_{sky}$ =  0.50 & 1.75 & 1.81 & 1.90 & 1.96 && 1.28 &  1.32 &  1.37 & 1.43 \\
 $f_{sky}$ = 0.75 & 1.45  &  1.51 & 1.57 & 1.61   && 1.10 & 1.12 &  1.16 & 1.20 \\
\hline
\end{tabular}
\caption{$n_s$ 1-$\sigma$ constraints in units of $10^{-3}$  from CMB and from CMB+BAO. ``CMB'' includes lensing. }
\label{table:n_s}
\end{table}

The best constraint for $n_s$ is 0.00145 from Table~\ref{table:n_s} for $10^6$ detectors, $1'$ beam, and 75\% $f_{sky}$. For reference, the constraints from TT, EE, TE spectra alone is 0.00151. This is a factor of five improvement from current \textit{Planck} best constraints (0.0073). We note that the gain over going to larger sky area plateaus once $f_{sky}$ hits 0.3 and adding BAO information helps by about 30\%. For $n_s$, the major input that changes the constraints is $f_{sky}$. In the CMB only case, going from $10^4$ to $10^5$ $N_{\rm det}$ and from $10^5$ to $10^6$ $N_{\rm det}$ improves the constraints by 5-10\% respectively. The improvement is better for smaller beam cases. Improvement from decreasing beam sizes per arc-minute is on the percent-level. 

Table~\ref{table:alp_s} summarizes the constraints for $\alpha_s$, the running of $n_s$, for experiments with $10^5$ and $10^6$ detectors, $1'-4'$ beams, covering 25-75\% of the sky, lensing included. The best constraint for $\alpha_s$ is 0.00160 (no lensing), 0.00146 (with lensing) and 0.00145 (lensing+BAO). Compared to the current best 1-$\sigma$ constraints from \textit{Planck} + \textit{WMAP} polarization + BAO of 0.009, future CMB experiments alone can give a factor of $5$ improvement.

\begin{table}[htbp]\centering
\ra{1.3}
\begin{tabular}{@{}lccccccccccl@{}}\hline
& \multicolumn{4}{c}{CMB} & \phantom{abc}&  \multicolumn{4}{c}{CMB+BAO}  \\
\cline{2-5} \cline{7-10} 
& $1'$&  $2'$  &  $3'$ & $4'$ &\phantom{abc}& $1'$ &  $2'$   &  $3'$  & $4'$   \\ \hline
$\mathbf{10^4}$ \textbf{detectors} \\
 $f_{sky}$ = 0.25  & 3.40 & 3.51 & 3.69 & 3.92 && 3.40 & 3.51 & 3.68 & 3.91 \\
 $f_{sky}$ =  0.50 & 2.58 & 2.66 & 2.79 & 2.96 && 2.58 & 2.66 & 2.78 & 2.95  \\
 $f_{sky}$ =  0.75 & 2.20 & 2.26 & 2.37 & 2.51 && 2.20 & 2.26 & 2.37 & 2.51 \\
$\mathbf{10^5}$ \textbf{detectors} \\
 $f_{sky}$ = 0.25  & 2.79 & 2.92 & 3.09 & 3.31 && 2.78 & 2.91 & 3.08 & 3.30  \\
 $f_{sky}$ =  0.50 & 2.09 & 2.17 & 2.29 & 2.45 && 2.08 & 2.17 & 2.29 & 2.44 \\
$f_{sky}$ =  0.75 & 1.76 &  1.83 & 1.93 & 2.06 && 1.76 & 1.82 & 1.92 & 2.05 \\
$\mathbf{10^6}$ \textbf{detectors} \\ 
 $f_{sky}$ = 0.25 & 2.32 &  2.46 & 2.65 & 2.87 && 2.27 & 2.42 & 2.63 & 2.86   \\
$f_{sky}$ =  0.50 & 1.75 & 1.83 & 1.96 & 2.12 && 1.72 &  1.82 & 1.96 & 2.11 \\
 $f_{sky}$ = 0.75 & 1.47 & 1.54 & 1.65 & 1.77 && 1.45 & 1.53 &  1.64 & 1.77 \\
\hline
\end{tabular}
\caption{$\alpha_s$ 1-$\sigma$ constraints in units of $10^{-3}$  from CMB and from CMB+BAO. ``CMB'' includes lensing. BAO measurements improve constraints in $\alpha$ by a few percent. The improvement is more significant for small sky fractions and small beam size scenarios. }
\label{table:alp_s}
\end{table}

Similar to $n_s$, once $f_{sky}$ reaches 0.3, the improvement on the constraints for $\alpha_s$ over adding sky area is quite flat. This is particularly true when lensing is included. We also note that the sensitivities and the beam sizes of the experiments do not have as big of an impact to the constraints as sky area does. So again, $f_{sky}$ is the key input. For $\alpha_s$, going from $10^4$ to $10^5$ $N_{\rm det}$ improves the constraints by 16-20\% and from $10^5$ to $10^6$ $N_{\rm det}$ by 13-17\%. Like $n_s$, the improvement is higher for smaller beams. With that noted, lensing has a larger impact on the constraints when beam sizes are smaller and when sky area is smaller. 

\subsection{Tensor-to-scalar ratio $r$}
\label{ssec:r}
Most of the sensitivity on $r$ comes from the B-mode polarization, which is already known to be much smaller in power than 
the E-modes and the temperature anisotropies. 
Since tensor modes are so far undetected, we set $r=0$ to be the fiducial value and study 
the 1-$\sigma$ uncertainty on $r$ for various experimental configurations. 
In this ``discovery'' phase, it is initially more advantageous to improve sensitivity on a few spatial modes \cite{jaffe2000}. 
When lensing starts to dominate ($r\sim 0.01$), high-resolution B-mode data can be used to reconstruct the lensing field with high 
fidelity for delensing. 
In this regime, there are complicated trade-offs between sky coverage, delensing residual, and polarized foregrounds.  
As a result, the optimized survey for $r$ can be significantly different from the lensing survey, which is important for constraining sum of neutrino mass and dark-energy equation of state. 
We use this separate section to study the strategies for the tensor search, taking delensing and foregrounds into account. This 
deep tensor survey is assumed to have equal observing time as the lensing survey, with enough frequency coverage to achieve the 
target foreground residual. 
If $r$ is large ($> 0.01$), CMB-S4 can be a powerful tool for detailed characterization of the tensor perturbations. 
In this section, we also look at how well $r$ and $n_t$ could be constrained for larger values of $r$. Another potentially interesting goal is to measure 
the 3-point functions of B-modes.  However, this is beyond the scope of this paper. 
 
\subsubsection{Method}
\label{sssec:r-method}
We use a smaller range of sky fraction when producing constraints for $r$ while maintaining the range of detector numbers and beam sizes. The grid of net sensitivities on $r$ is given in Table  \ref{table:tensorsensifsky}. 
The range of $f_{sky}$ is chosen to observe the so-called ``recombination bump'' at multipole $\ell\sim 100$. 
The inflationary tensor signal at very large
angular scales (the ``reionization bump'' at $\ell\sim 8$) can in principle exceed lensing even for low values of
$r$ ($r < 10^{-3}$). It is buried deep in galactic foregrounds but could in principle be recoverable with frequency-component
separation. We acknowledge this exciting possibility but choose not to include the reionization bump in the forecast for simplicity.

\begin{center}
\begin{table}[ht]\centering\footnotesize
\ra{1.3}
\begin{tabular}{@{}r|p{0.1cm}p{0.875cm}p{0.875cm}p{0.875cm}p{0.875cm}p{0.875cm}p{0.875cm}p{0.875cm}@{}}\hline
$N_{\rm det}$ $\backslash$ f$_{\text{sky}}$ && 0.0004& 0.0006 & 0.0025 & 0.01 & 0.0225 & 0.04 & 0.25 \\ \hline
10,000 \ \ && 0.13 & 0.17 & 0.33 & 0.67 & 1.00 & 1.34 & 3.34  \\
20,000 \ \ && 0.09 & 0.12 & 0.24 & 0.47 & 0.71 & 0.95  & 2.37 \\
50,000 \ \ && 0.06 & 0.07 & 0.15 & 0.30 & 0.45 & 0.60 & 1.50  \\
100,000 \ \ && 0.04 & 0.05 & 0.11 & 0.21 & 0.32 & 0.42 & 1.06 \\
200,000 \ \  && 0.03 & 0.04 & 0.07  & 0.15 & 0.22 & 0.30 & 0.75 \\
500,000 \ \ && 0.02 & 0.02 & 0.05 & 0.09 & 0.14 & 0.19 & 0.47 \\
1,000,000 \ \  && 0.013 & 0.02 & 0.03 & 0.07 & 0.10 & 0.13 & 0.33 \\
\hline
\end{tabular}
\caption{Table of experimental sensitivities on $r$ given detector count and sky coverage, in $\mu$K-arcmin, used for investigating the constraints for the tensor-to-scalar ratio}
\label{table:tensorsensifsky}
\end{table}
\end{center}

At $r=0.01$, the magnitude of the primordial gravitational wave B-mode power is comparable to lensing at $\ell = 100$. 
Lensing-induced $B$ modes have the same frequency dependence as the CMB and cannot be distinguished by
multi-frequency observations.
The lensing contamination can be debiased (i.e., subtracted in power-spectrum space) in the same
way instrumental noise is removed in temperature power spectrum measurements when the expected lensing power is well known.
Alternatively, the lensing deflection field can be reconstructed from arcminute-scale $B$-mode measurements, and the
expected lensing contamination to degree-scale $B$ modes predicted and subtracted
from the observed $B$-mode map.  The quadratic estimators \cite{2002ApJ...574..566H} and maximum-likelihood estimators \cite{Hirata:2003ka} are techniques that have been developed to delens CMB maps. Here we use the method outlined in 
\citet{2012JCAP...06..014S} to forecast the residual noise level after iterative delensing, which converges to 
the maximum-likelihood solution. 

Polarized synchrotron and thermal dust constitute the major sources of astrophysical foregrounds in the measurement of the B-mode power spectrum. In this study, we use the Planck Sky Model \cite{2013A&A...553A..96D} to estimate the level of polarized foreground we would have given that we observe the cleanest patches of the stated sky area. We compute the BB spectrum from polarized dust and synchrotron at $95$ GHz, which is amongst the cleanest frequencies for small sky areas \cite{2009AIPC.1141..222D} and expect them to be cleaned to 1-10\% \cite{Clive2013PrivateComm} of the observed foreground BB power using existing component separation techniques \cite{2013arXiv1303.5072P}. For synchrotron, we use a curved power law for emission with the reference frequency set at $20$ GHz and curvature at -0.3. We use the model developed in \citet{2008A&A...490.1093M} for the synchrotron emission index. For polarized dust, we employ the galactic polarization model from the same paper \cite{2008A&A...490.1093M}, where the thermal dust emission in intensity is based on Model 7 of \citet{1999ApJ...524..867F}, with the mean polarization fraction set to ~5\%. We then incorporate the cleaned foreground level as a noise term in the Fisher forecast.

The effect of tensor modes on the temperature and E-mode polarization power spectra is small compared to that on the B-mode power spectrum. 
So we can forecast constraints on $r$ using only the B-mode spectrum to obtain almost all of the constraining power. In the B-mode spectrum, given perfect delensing and aside from the B-power enhancement at low-ell by $\tau$ during reionization, the constraint on $r$ is independent of the rest of the $\nu\Lambda CDM$ parameter space. Therefore, in the forecast for $r$, losing the extra constraining power from $\tau$, the Fisher matrix $F_{ij}$ in Eq.~\eqref{eq:Fij_def} can be reduced to:

\begin{equation}
F_{rr} = \sum_\ell \frac{(2\ell+1)f_{sky}}{2} \frac{1}{(\delta C_\ell)^2} \left(\frac{\partial C^{B_{tens}}_\ell}{\partial r}\right)^2
\label{eq:Frr}
\end{equation}

with
\begin{equation}
\delta C_\ell = C^{B_{tens}}_\ell + N^{BB}_\ell + N^{fg}_\ell + N^{res}_\ell
\end{equation}
where $C^{B_{tens}}$ is the tensor contribution of the BB power spectrum, $N^{BB}_\ell$ is defined in Eq.~\eqref{eq:beamnoise}, $N^{fg}_\ell$ is noise from foreground, and $N^{res}_\ell$ is the BB residual after delensing the B-mode power spectrum using the fast algorithm developed in Appendices A and B of~\citet{2012JCAP...06..014S}. 

In the scenario that delensing is imperfect, the constraints on $r$ does depend on our knowledge of other parameters that determine the amplitude and shape of the lensing power spectrum. This is because the uncertainty in the lensing power spectrum cascades into the uncertainty in $N^{res}_\ell$. Given a CMB-S4 like experiment with $5\times10^5$ detectors, $3' $ beam, and $f_{sky} = 0.5$, we 
estimate a $0.5\%$ uncertainty (95\% C.L.) in the level of BB lensing amplitude averaged over $\ell$ range of 2 to 3000. 
If the amplitude of $N^{res}_\ell$ were $10\%$ of the lensing BB amplitude, the uncertainty in $N^{res}_\ell$ would be $0.05\%$, which is small compared to other terms.

Since we take the fiducial $r$ to be 0, the constraints in $r$ inform us how well (in $\sigma$) we can differentiate $r\ne0$ from $r = 0$. For the specific experimental cases considered in this work, a comparison of Eq.~\eqref{eq:Frr} with the formalism developed in Ref.~\cite{2011PhRvD..84f3005E} shows a relative difference of $ \leq 1\,\%$.

\subsubsection{Results and discussion}
Table \ref{table:r_constraints} presents the 1-$\sigma$ constraints for r for a range of detector counts, sky fractions, beam sizes, and foreground residuals.
The main result is that almost all of these experiments can constrain $r$ to 1 $\sigma$ at 0.002 or better. This means that we can certainly determine whether the field range is sub-Planckian or super-Planckian if $N_e$ = 30, and it will drive theoretical research on models of inflation. 

\begin{table}[htbp]\centering\footnotesize
\ra{1.3}
\begin{tabular}{@{}lccccccccccl@{}}\hline
& \multicolumn{4}{c}{1\% foreground} & &  \multicolumn{4}{c}{10\% foreground}  \\
\cline{2-5} \cline{7-10} 
& $1'$&  $2'$  &  $3'$ & $4'$ && $1'$ &  $2'$   &  $3'$  & $4'$   \\ \hline
$\mathbf{10^4}$ \textbf{detectors} \\
$f_{sky}$ =  0.0025  & 0.050 & 0.053 & 0.059 & 0.066 && 0.176 & 0.182 & 0.191 & 0.203  \\
 $f_{sky}$ =  0.04 & 0.052 & 0.054 & 0.056 & 0.058 && 0.119 & 0.121 & 0.124 & 0.128 \\
$f_{sky}$ =  0.25 & 0.072 & 0.072 & 0.073 & 0.074 && 0.146 & 0.147 & 0.148 & 0.150 \\
$\mathbf{10^5}$ \textbf{detectors} \\
 $f_{sky}$ = 0.0025  & 0.023 & 0.024 & 0.028 & 0.032 && 0.120 & 0.125 & 0.134 & 0.145  \\
 $f_{sky}$ =  0.04 & 0.018 & 0.019 & 0.020 & 0.022 && 0.066 & 0.068 & 0.071 & 0.074 \\
$f_{sky}$ =  0.25 & 0.022 & 0.022 & 0.023 & 0.024 && 0.072 & 0.073 & 0.075 & 0.077 \\
$\mathbf{10^6}$ \textbf{detectors} \\ 
$f_{sky}$ = 0.0025 & 0.015 & 0.016 & 0.018 & 0.021 && 0.096 & 0.101 & 0.108 & 0.119   \\
$f_{sky}$ =  0.04 & 0.009 & 0.010 & 0.011 & 0.012 && 0.048 &  0.049 & 0.052 & 0.056 \\
 $f_{sky}$ = 0.25 & 0.010 & 0.011 & 0.012 & 0.013 && 0.051 & 0.052 &  0.054 & 0.056 \\
\hline
\end{tabular}
\caption{1-$\sigma$ constraints on $r$, in units of $10^{-2}$. For $1\%$ foreground residuals, all the constraints are well below $2 \times 10^{-3}$ in the wide range of experimental configurations we considered. Even for 10 \% foregrounds in map space, all except one of the values listed here are below $2 \times 10^{-3}$.}
\label{table:r_constraints}
\end{table}

\begin{figure}[htbp]
\includegraphics[width=\linewidth]{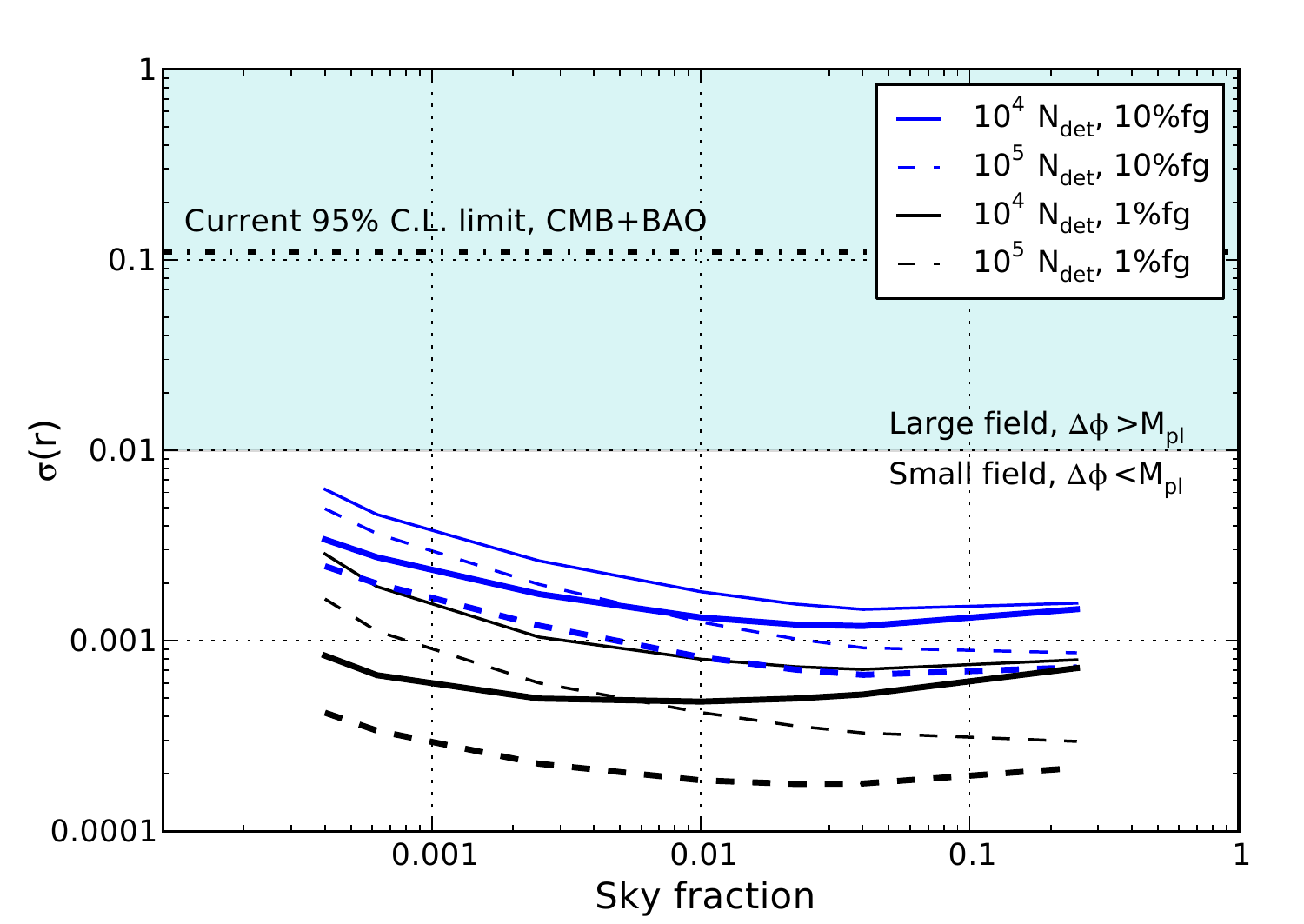}
\caption{This plot shows 1 $\sigma$ constraints in $r$ vs. sky fraction for a few experimental configurations, including combinations for $1'$ (thick lines) and $8'$ (thin lines) beams, and $10^4$ and $10^5$ detectors. Two levels of foreground residuals are shown. These are a few examples of a broad class of experimental configurations that yield constraints that are orders of magnitude smaller than $r = 0.01$, which could differentiate super-Planckian and sub-Planckian inflation field range. }
\label{fig:r_fsky}
\end{figure}

Fig.~\ref{fig:r_fsky} shows the 1-$\sigma$ constraint on $r$ as a function of sky fraction for a few experimental configurations. We pick the two ends of beam sizes, $1'$ and $8'$, and medium and low end of detector number, $10^4$ and $10^5$, to illustrate the reach of CMB-S4 in $r$.
Under a wide range of experimental parameters, we find a broad minimum in $\sigma(r)$ to lie between $0.01<f_{sky}<0.1$.
Even for the most ambitious configuration we studied ($10^6$ detectors, $1'$ beam), we find the optimal survey area to be around
1,000 square degrees.

\begin{figure*}[!ht]
\includegraphics[width=\linewidth]{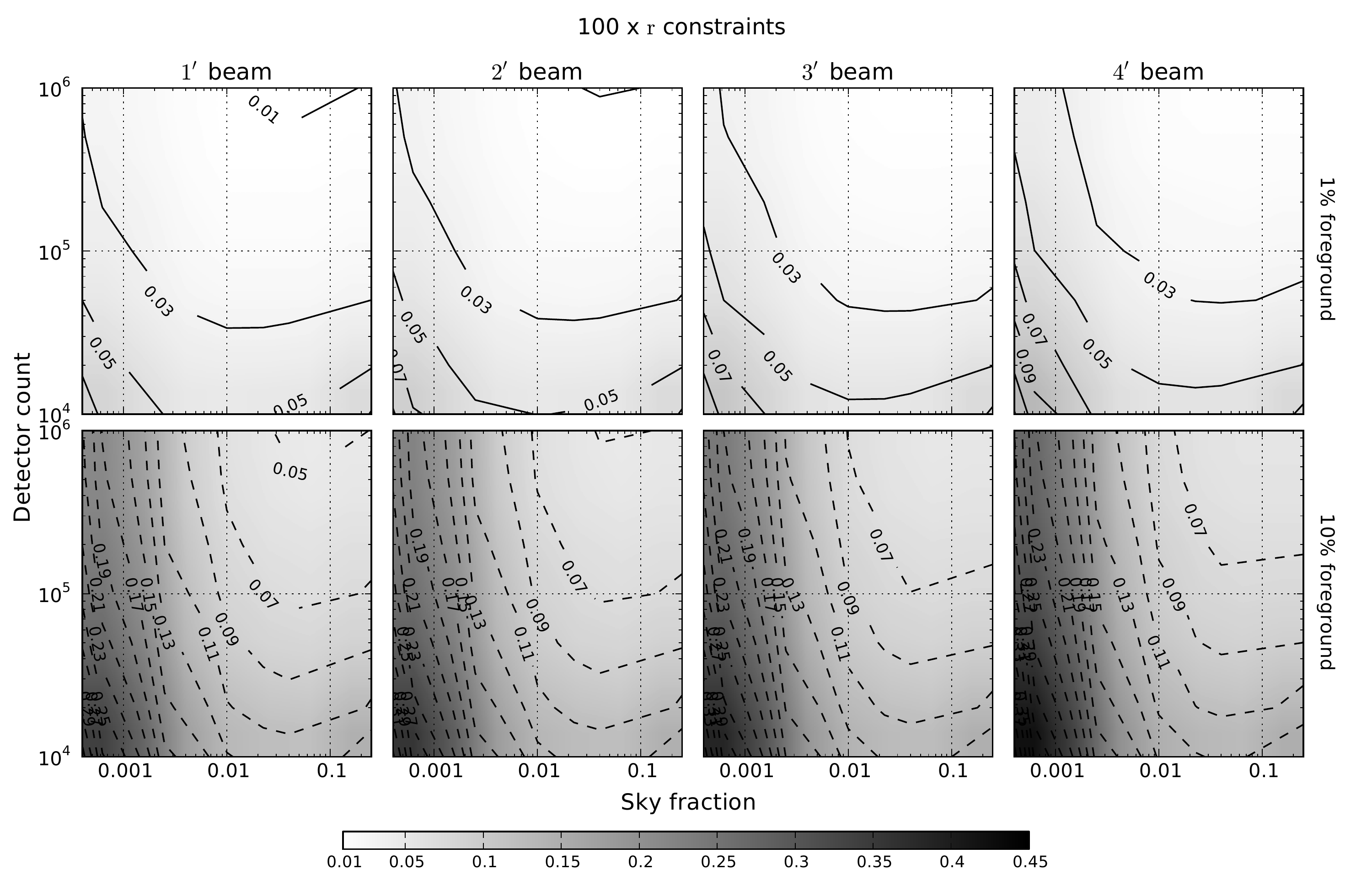}
\caption{This panel shows the 1 $\sigma$ constraints of $r$, in units of $10^{-2}$, as a function of detector number and sky fraction. The beam sizes of the experiments shown from left to right go from $1'-4'$. The top and bottom row shows constraints with 1\% and 10\% foreground residuals in power respectively. The striking feature from this set of plots is that aside from small left hand strips of the 10\% foreground plots, all the constraints are $< \,0.002 \, (0.01 \times 0.2)$. In other words, there are many experimental configurations that could tell $r = 0.01$ from $r=0$ by 5$\sigma$, thus differentiating $\Delta \phi$ is super-Planckian or sub-Planckian. This set of plots also highlights the difference foreground removal makes in the constraints for $r$. }
\label{fig:r_panel}
\end{figure*}

Fig.~\ref{fig:r_panel} shows the constraints on $r$ as a function of detector count and sky coverage for $1'-4'$ beams with 1\% and 10\% foreground residuals in power. The constraints with 0.25\% and 1\% foreground residuals are 3-6 times better than those with 10\% foregrounds. But even with 10\% foreground, if one covers more than 0.2\% of the sky with a small beam, $\sigma(r) < 0.002$ is still achievable. 

We compared iterative and quadratic delensing methods and learned that iterative delensing help constrain $r$ a lot more than quadratic delensing at small sky area compared to large sky area. For example, it improves the constraints at smaller patches by a factor of 6 to 20 for $f_{sky} = 0.01$ for $10^5$ detectors depending on what beam sizes the experiment has. However, the improvement is only a factor of less than 2 for $f_{sky} = 0.25$. This is important at the discovery phase and when we do not observe  large patches.

Beam size matters less when sky area is big because having a larger sky area provides more modes that debias the result regardless of how well delensing is done. On the other hand, for smaller sky fraction, one relies more heavily on delensing to provide competitive constraints, thus a smaller beam is advantageous. However, this improvement from delensing using a small beam gets greatly washed out if the foreground level is high, especially when the experiment has high sensitivity. For example, at $f_{sky}$ = 0.04, for different beam sizes and detector number, at least a factor of 2 and at most an order of magnitude constraining power to $r$ is lost when foreground residual is 10\% and not 0.25\% (a very foreground clean case). 

\textit{$n_t$ and $r$ constraints if $r > 0.01 $ --}
In the scenario that $r$ is bigger than 0.01, the B-mode peak around $\ell$ of 100 is higher than the lensing B-mode power. We can therefore constrain $r$ and $n_t$ without delensing. We run the forecast using only $C^{BB}_{\ell}$ for $\ell$-range of 10 to 500 for experiments with a grid shown in Table~\ref{table:sensifsky} and $4'$ beam. The reason for this $\ell$ range is to exclude the reionization bump at $\ell < 10$ and noise above $\ell > 500$. Beam size is not critical here because we are not delensing.  We picked a few fiducial $r$ values bigger than 0.02 to illustrate the trend. The fiducial $n_t$ value is $n_t=0$, i.e. we are not imposing the consistency relation between $r$ and $n_t$ valid in single-field slow-roll inflation. 
Table~\ref{table:r_nt} presents the constraints for $r$ and $n_t$ for several fiducial $r$ values.

\begin{table}[ht]\centering
\ra{1.3}
\begin{tabular}{@{}lccccccccccl@{}}\hline
& \multicolumn{3}{c}{$\sigma(r)$} & \phantom{abc}& \multicolumn{3}{c}{$\sigma(n_t)$} \\
\cline{2-4} \cline{6-8}
$f_{sky}$ & 0.25 & 0.50 & 0.75 &\phantom{abc} & 0.25 & 0.50 & 0.75 \\ \hline
$\mathbf{r_{fid} = 0.2}$ \\
$10^4 N_{\rm det}$ & 0.030 & 0.026 & 0.025 && 0.09 & 0.08 & 0.07 \\
$10^5 N_{\rm det}$ & 0.021 & 0.016 & 0.014 && 0.07 & 0.05 & 0.04 \\
$10^6 N_{\rm det}$ & 0.020 & 0.015 & 0.012 && 0.06 & 0.05 & 0.04 \\
$\mathbf{r_{fid} = 0.1}$ \\
$10^4 N_{\rm det}$  & 0.022 & 0.020 & 0.019 && 0.13 &  0.11 & 0.11 \\
$10^5 N_{\rm det}$  & 0.016 & 0.012 & 0.010 &&  0.10 & 0.07 & 0.06 \\
$10^6 N_{\rm det}$ & 0.015 & 0.011 & 0.009 &&  0.09 &  0.07 & 0.05 \\
$\mathbf{r_{fid} = 0.05}$ \\
$10^4 N_{\rm det}$ & 0.016 & 0.015 & 0.014 &&  0.18 & 0.16 & 0.16 \\
$10^5 N_{\rm det}$ & 0.012 & 0.009 & 0.007 &&  0.14 & 0.10 & 0.08 \\
$10^6 N_{\rm det}$  & 0.011& 0.008 & 0.007 &&  0.13 & 0.09 & 0.08 \\
$\mathbf{r_{fid} = 0.02}$ \\
$10^4 N_{\rm det}$ & 0.012 & 0.011 & 0.011 &&  0.31 &  0.29 & 0.28 \\
$10^5 N_{\rm det}$ & 0.008 & 0.006 & 0.005 &&  0.22 &  0.16 & 0.14 \\
$10^6 N_{\rm det}$ & 0.008 & 0.005 & 0.004 &&  0.21 &  0.15 & 0.12 \\
\hline
\end{tabular}
\caption{1-$\sigma$ constraints for $r$ and $n_t$ for various detector count and sky fraction at $4'$ beam size.}
\label{table:r_nt}
\end{table}

We observe that for the few high fiducial $r$ values listed, we can have $3\sigma$ to $5\sigma$ constraints without delensing in the grid we consider in this work. However, with this approach, $n_t$ cannot be measured to a precision that allows us to verify the consistency relation between $r$ and $n_t$ in single-field slow-roll inflation.

\section{Conclusion}
\label{sec:conclusion}
In this work, we forecasted how well a highly capable next-generation ground-based CMB experiment can constrain cosmological parameters of fundamental physics relevant to both high energy physics and cosmology. We forecast for a range of experimental inputs -- $10^4$ to $10^6$ detectors, $1'-4'$ beams ($6'$, $8'$ in some cases), $1-75\%$ sky fraction in order to see how the constraints for each parameters vary with these inputs. 

We detailed in section~\ref{sec:methods} the methods used to estimate the performance of a given CMB experimental design. We presented our results in sections~\ref{sec:neutrinos},~\ref{sec:DarkEnergy}, ~\ref{sec:DarkMatter}, and ~\ref{sec:Inflation}. 
Here we quote the range of constraints each parameter falls in for experiments with $10^4-10^6$ detectors, $1'$ to $4'$ beam, and 25\% to 75\% $f_{sky}$ as illustrations and summarize the CMB only parameter improvement dependence on $N_{\rm det}$, beam size, and $f_{sky}$:
\begin{itemize}
         \item $ 0.0156  \leq \sigma(N_{\rm eff}) \leq 0.0690$ (CMB): \\
	increasing $N_{\rm det}$ from $10^4$ to $10^5$ improves $\sigma(N_{\rm eff})$ by about 30\%, same for going from $10^5$ to $10^6$; decreasing beam size improves $\sigma(N_{\rm eff})$ by $\sim$10\% per arc-minute; $\sigma(N_{\rm eff})$ is not sample variance limited. 

	\item $ 15  \leq \sigma(M_{\nu}) \leq 24 $ [meV] (CMB+BAO): \\
	increasing $N_{\rm det}$ from $10^4$ to $10^5$ improves $ \sigma(M_{\nu})$ by 10-20\% (smaller beam gives better improvement with $N_{\rm det}$), while increasing $N_{\rm det}$ from $10^5$ to $10^6$ improves $\sigma(M_{\nu})$ by 5-15\%; decreasing beam size improves $\sigma(M_{\nu})$ at percent-levels per arc-minute; $\sigma(M_{\nu})$ is sample variance limited.

	\item $164 \leq$ DETF-FoM $\leq 303$ (CMB+BAO+$H_0$):\\
	increasing $N_{\rm det}$ from $10^4$ to $10^5$ improves FoM by more than factor of 2, same for going from $10^5$ to $10^6$; decreasing beam size improves the FoM by a few \% to 10s of \% depending on configuration; FoM is not sample variance limited.
	
	\item $0.00588 \leq \sigma(p_{ann}) \leq 0.0110\, [3\times10^{-26}$ cm$^3$/s/GeV] (CMB):\\
	increasing $N_{\rm det}$ from $10^4$ to $10^5$ improves $\sigma(p_{ann})$ by about 4\%, same for going from $10^5$ to $10^6$; decreasing beam size improves $\sigma(p_{ann})$ by $\lesssim$ 1\%; $\sigma(p_{ann})$ is sample variance limited.

	\item $0.00074 \leq \sigma(\Omega_K) \leq 0.0014 $ (CMB+BAO+$H_0$):\\
	increasing $N_{\rm det}$ from $10^4$ to $10^5$ improves $\sigma(\Omega_K)$ by about 20\%, while increasing $N_{\rm det}$ from $10^5$ to $10^6$ improves $\sigma(\Omega_K)$ by 10-20\% (smaller beam gives a better improvement with $N_{\rm det}$); decreasing beam size improves $\sigma(\Omega_K)$ at percent levels per arc-minute; $\sigma(\Omega_K)$ is sample variance limited. 
	
	\item $0.00110 \leq \sigma(n_s) \leq 0.00236$ (CMB+BAO): \\
	increasing $N_{\rm det}$ from $10^4$ to $10^5$ improves $\sigma(n_s)$ by $\sim\, 5-10\%$ (smaller beam gives a better improvement with $N_{\rm det}$), same for going from $10^5$ to $10^6$; decreasing beam size improves $\sigma(n_s)$ at percent levels per arc-minute; $\sigma(n_s)$ is sample variance limited. 
	 
	\item $0.00145 \leq \sigma(\alpha_s) \leq 0.00330$ (CMB+BAO): \\
	increasing $N_{\rm det}$ from $10^4$ to $10^5$ improves $\sigma(\alpha_s)$ by 16-20\%, and by 13-17\% going from $10^5$ to $10^6$ (smaller beam gives a better improvement with $N_{\rm det}$); decreasing beam size improves $\sigma(\alpha_s)$ at percent levels per arc-minute; $\sigma(\alpha_s)$ is sample variance limited. 
	 
	\item $0.00009 \leq \sigma(r) \leq 0.00203$ for 1\% and 10\% foreground residual: \\
	$\sigma(r)$ is foreground limited; when foreground is high, the optimal $f_{sky}$ shifts higher; decreasing beam size improves $\sigma(r)$ at percent levels per arc-minute (the slope of this trend with beam increases with lower sky coverage). 
	
\end{itemize}
Detailed constraints in specific cases can be read off from tables listed in each section.\\

Besides learning the approximate ranges of how well these parameters can be constrained with CMB-S4, we also learn how the constraints improve as functions of $N_{\rm det}$, beam size, and $f_{sky}$: 
\begin{itemize}
    \item For all parameters, except $r$, increasing $f_{sky}$ always improve the constraints even though the overall sensitivity of the experiment decreases. 
    \item For all parameters, except those related to dark-energy equation of state, going from $10^5$ to $10^6$ detectors yields the same or less percentage improvement on the constraints than going from $10^4$ to $10^5$ detectors. The improvements range from a few to tens of percent. For $M_{\nu}$, the improvement beyond $10^5$ detectors is marginal when BAO signal is added.
    \item Dependence on beam size is quite mild -- constraints on $N_{\rm eff}$ improves by about 10\% per arc-minute decrease while for all other parameters it is around a few \% improvement per arc-minute decrease. 
\end{itemize}

We envision CMB-S4 to be a powerful next-generation ground-based CMB polarization experiment with high-resolution and high-sensitivity. With CMB-S4, we showed that most constraints on cosmological parameters are sample variance limited.
Combining these data sets with space-borne observations will allow access to larger sky fraction, thus further improving the constraints on sample variance limited parameters.

\acknowledgements{
This work and collaboration were encouraged by the preparation for the Snowmass Community Summer Studies 2013. We thank Bradford Benson, John Carlstrom, Tom Crawford, Daniel Eisenstein, Daniel Green, Ryan Keisler, John Kovac, Lloyd Knox, Eric Linder, Toshiya Namikawa, Peter Redl, Leonardo Senatore, Neelima Sehgal, and Kendrick Smith for useful discussions. WLKW thanks Olivier Dor\'{e} for assistance with the Fisher matrix code. 

We acknowledge the use of the PSM, developed by the Component Separation Working Group (WG2) of the \textit{Planck} Collaboration. We also used CAMB and an implementation of iterative delensing developed by Wei-Hsieng Teng. 
CD was supported by the National Science Foundation grant number AST-0807444, NSF grant number PHY-088855425, and the Raymond and Beverly Sackler Funds.
CLK acknowledges the support of an Alfred P. Sloan Research Fellowship and an NSF Faculty Early Career Development (CAREER) Award (award number: 1056465). 
ATL acknowledges support from the U.S. Department of Energy Office of Science.
OZ acknowledges support by National Science Foundation through grants ANT-0638937 and ANT-0130612. }

\onecolumngrid
\appendix*
\section{DETF constraints}

\begin{table*}[htbp]\footnotesize\centering
\ra{1.3}
\begin{tabular}{@{}lccccccccccccccccl@{}}\hline
& \multicolumn{5}{c}{CMB} & \phantom{abc}&  \multicolumn{5}{c}{CMB+BAO} & \phantom{abc}& \multicolumn{5}{c}{CMB+BAO+1\% H$_0$ prior} \\
\cline{2-6} \cline{8-12} \cline{14-18} 
&{FoM} & {$\sigma(w_0)$}  & {$\sigma(w_a)$} &{$\sigma(w_p)$} & {$\sigma(\Omega_K)$} &\phantom{abc} & {FoM} & {$\sigma(w_0)$}  & {$\sigma(w_a)$} &{$\sigma(w_p)$} & {$\sigma(\Omega_K)$} &\phantom{abc} & {FoM} & {$\sigma(w_0)$}  & {$\sigma(w_a)$} &{$\sigma(w_p)$} & {$\sigma(\Omega_K)$} \\ \hline
$\mathbf{10^4}$ \textbf{detectors} \\
0.25, $1'$  & 8 & 0.26 & 1.2 & 0.10 & 12.0 && 86 & 0.12 & 0.39 & 0.030 & 1.3 && 170 & 0.069 & 0.26 & 0.023 & 1.3 \\
 0.50, $1'$ & 12 & 0.21 & 0.97 & 0.086 & 9.4 && 100 & 0.11 & 0.36 & 0.028 & 1.1 && 187 & 0.067 & 0.24 & 0.022 &  1.0 \\
 0.75, $1'$ & 15 & 0.18 & 0.84 & 0.078 & 7.9 && 108 & 0.11 & 0.34 & 0.027 & 1.0 && 196 & 0.065 & 0.24 & 0.021 & 0.9 \\[0.1cm]
 
 0.25, $2'$ & 7 & 0.28 & 1.3 & 0.11 & 12.4  && 84 & 0.13 & 0.40 & 0.030 & 1.4 && 169 & 0.069 & 0.26 & 0.023 & 1.3 \\
 0.50, $2'$ & 11 & 0.22 & 1.0 & 0.093 & 9.7 && 97 & 0.12 & 0.37 & 0.028 & 1.1 && 185 & 0.067 & 0.25 & 0.022 &  1.0 \\
 0.75, $2'$ & 14 & 0.19 & 0.88 & 0.084 & 8.2 && 104 &  0.11& 0.35 & 0.027 & 1.0 && 193 & 0.066 & 0.24 & 0.022 & 0.9 \\[0.1cm]
 
 0.25, $3'$ & 6 & 0.29  &1.3  & 0.12 & 12.8  && 81 & 0.13 & 0.40 & 0.031 & 1.4 && 166 & 0.070 & 0.26 & 0.023 & 1.3  \\
 0.50, $3'$ & 9 &  0.23  & 1.1 & 0.10 & 10.1 && 93 & 0.12 & 0.38 & 0.029 & 1.1 &&  181 & 0.068 & 0.25 & 0.022 & 1.0 \\
 0.75, $3'$ & 12 & 0.21  & 0.93 & 0.093 & 8.5 && 99 & 0.12 & 0.36 & 0.028 & 1.0 &&  189 & 0.066 & 0.24 & 0.022 & 0.9 \\[0.1cm]
 
 0.25, $4'$ & 5 & 0.32 & 1.4 & 0.14 & 13.3 && 78 & 0.13 & 0.41 & 0.031 & 1.4 && 164 & 0.070 & 0.26 & 0.023 & 1.4 \\
 0.50, $4'$ & 8 & 0.25 & 1.1 & 0.12 &  10.5 && 88 & 0.13 & 0.39 & 0.029 & 1.1 &&  178 & 0.068 & 0.25 & 0.023 & 1.1 \\
 0.75, $4'$ & 10 & 0.22 & 0.98 & 0.11 & 8.8 && 94 & 0.12 & 0.38 & 0.028 & 1.0 &&  185 & 0.067 & 0.24 & 0.022 & 1.0 \\

$\mathbf{10^5}$ \textbf{detectors} \\
 0.25, $1'$  &18 & 0.18 & 0.80 & 0.069 & 9.3  && 106 & 0.11 & 0.34 & 0.028  & 1.2 && 187 & 0.066 & 0.24 & 0.022 & 1.2 \\
 0.50, $1'$ & 28 & 0.14 & 0.66 & 0.054  & 7.2  && 132 & 0.093 & 0.30 & 0.025 & 1.0 && 214 & 0.062 & 0.23 & 0.020 & 0.9 \\
 0.75, $1'$ & 37 & 0.13 & 0.57 & 0.047 & 6.0  && 153 & 0.085 & 0.28 & 0.024 & 0.8 && 234 & 0.060 & 0.22 & 0.020 & 0.8 \\[0.1cm]
 
 0.25, $2'$ & 15 & 0.19 & 0.88 & 0.074 & 9.9 && 101 & 0.11 & 0.35 & 0.028 &  1.3 && 183 & 0.067 & 0.25 & 0.022 & 1.2 \\
 0.50, $2'$ & 25 & 0.15 & 0.70 & 0.058 & 7.4 && 126 & 0.097 & 0.31 & 0.026 & 1.0 && 209 & 0.063 & 0.23 & 0.021 & 0.9 \\
 0.75, $2'$ & 33 & 0.13 & 0.60 & 0.050 & 6.1 && 144 & 0.089 & 0.29 & 0.024 & 0.9 && 227 & 0.061 & 0.22 & 0.020 & 0.8 \\[0.1cm]
 
 0.25, $3'$ & 13 & 0.21 & 0.97 & 0.081 & 10.3 && 96 & 0.12 & 0.36 & 0.029 & 1.3 && 179  & 0.067 & 0.25 & 0.022 & 1.2 \\
 0.50, $3'$ & 21 & 0.16 & 0.76 & 0.064 & 7.7  && 118 & 0.10 & 0.32 & 0.026 & 1.1 && 203 & 0.064 & 0.23 & 0.021 & 1.0 \\
 0.75, $3'$ & 28 & 0.14 & 0.64 & 0.056 & 6.3  && 134 & 0.094 & 0.30 & 0.025 & 1.0 && 219 & 0.062 & 0.22 & 0.020 & 0.8 \\ [0.1cm]
 
 0.25, $4'$ & 10 & 0.24 & 1.1 & 0.092 & 10.7 && 90 & 0.12 & 0.38 & 0.029  & 1.3 && 175 & 0.068 & 0.25 & 0.023 & 1.3 \\
 0.50, $4'$ & 17 & 0.18  & 0.82 & 0.073 & 8.0 && 110 & 0.11 & 0.34 & 0.027 & 1.0 && 197 & 0.065 & 0.24 & 0.021 & 1.0 \\
 0.75, $4'$ & 23 & 0.15 & 0.69 & 0.064 & 6.5  && 124 & 0.10 & 0.31 & 0.026 & 0.9 && 211 & 0.063 & 0.23 & 0.021 & 0.9 \\
 
$\mathbf{10^6}$ \textbf{detectors} \\ 
 0.25, $1'$  & 50 & 0.10 & 0.38 & 0.052 & 6.3 && 146 & 0.085 & 0.27 & 0.026 & 1.0 && 227 & 0.058 & 0.21 & 0.021 & 1.0 \\
 0.50, $1'$ & 74 & 0.085 & 0.34 & 0.040 & 5.0 && 190 & 0.072 & 0.23 & 0.023 & 0.8 && 270 & 0.054 & 0.20 & 0.019 & 0.8 \\
 0.75, $1'$ & 92 & 0.077 & 0.32 & 0.034 & 4.3 && 225 & 0.065 & 0.21 & 0.021 & 0.7 &&  303 & 0.051 & 0.19 & 0.018 & 0.7 \\[0.1cm]
 
 0.25, $2'$ & 40 & 0.11 & 0.45  & 0.056 & 7.0  && 133 & 0.090 & 0.29 & 0.026 & 1.1 && 214 & 0.060 & 0.22 & 0.021 & 1.1 \\
 0.50, $2'$ & 59 & 0.096 & 0.40 & 0.042 & 5.5 && 173 & 0.077 & 0.25 & 0.023 & 0.9 && 253 & 0.056 & 0.20 & 0.019 & 0.9 \\
 0.75, $2'$ & 75 & 0.086 & 0.37 & 0.036 & 4.8 && 204 & 0.069 & 0.23 & 0.021 & 0.8 && 283 & 0.053 & 0.19 & 0.018 & 0.8 \\[0.1cm]
 
 0.25, $3'$ & 30 & 0.13 & 0.55 & 0.061 & 8.0  && 120 & 0.097 & 0.31 & 0.027 & 1.2 && 202  & 0.063 & 0.23 & 0.021 & 1.2 \\
 0.50, $3'$ & 45 & 0.11 & 0.48 & 0.047 & 6.2  && 154 & 0.083 & 0.27 & 0.024 & 0.9 &&  235 & 0.059 & 0.21 & 0.020 & 0.9 \\
 0.75, $3'$ & 58 & 0.099 & 0.43 & 0.040 & 5.2  && 182 & 0.075 & 0.25 & 0.022 & 0.8 && 261 & 0.056 & 0.20 & 0.019 & 0.8 \\[0.1cm]
 
 0.25, $4'$ & 22 & 0.16 & 0.67 & 0.069 & 8.8  && 109 & 0.10 & 0.33 & 0.028 & 1.3 && 191 & 0.065 & 0.24 & 0.022 & 1.2 \\
 0.50, $4'$ & 34 & 0.13 & 0.56 & 0.053 & 6.7  && 138 & 0.090 & 0.29 & 0.025 & 1.0 && 221 & 0.061 & 0.22 & 0.020 & 0.9 \\
 0.75, $4'$ & 44 & 0.11 & 0.50 & 0.045 & 5.5  && 161 & 0.082 & 0.27 & 0.023 & 0.9 && 243 & 0.058 & 0.21 & 0.019 & 0.8 \\
\hline
\end{tabular}
\caption{The constraints for $w_0$, $w_a$, $w_p$, and $\Omega_K$, and the FoM for various experimental setups using CMB (with lensing), and adding external information like DESI BAO and a $1\%$ $H_0$ prior. The first column contains ($f_{sky}$, beam size) parameters. The constraints for $\Omega_K$ are in units of $10^{-3}$. }
\label{table:DEparams}
\end{table*}

\begin{table*}[htbp]\footnotesize\centering
\ra{1.3}
\begin{tabular}{@{}lccccccccccccccccl@{}}\hline
& \multicolumn{5}{c}{CMB} & \phantom{abc}&  \multicolumn{5}{c}{CMB+BAO} & \phantom{abc}& \multicolumn{5}{c}{CMB+BAO+1\% H$_0$ prior} \\
\cline{2-6} \cline{8-12} \cline{14-18} 
&{FoM} & {$\sigma(w_0)$}  & {$\sigma(w_a)$} &{$\sigma(w_p)$} & {$\sigma(\Omega_K)$} &\phantom{abc} & {FoM} & {$\sigma(w_0)$}  & {$\sigma(w_a)$} &{$\sigma(w_p)$} & {$\sigma(\Omega_K)$} &\phantom{abc} & {FoM} & {$\sigma(w_0)$}  & {$\sigma(w_a)$} &{$\sigma(w_p)$} & {$\sigma(\Omega_K)$} \\ \hline
$\mathbf{10^4}$ \textbf{detectors} \\
 0.25, $1'$  & 9 & 0.26 & 1.2  &  0.096 & 6.1  && 160  & 0.10 & 0.28  & 0.022  &  1.2 &&  299 & 0.059 & 0.18  & 0.018  &  1.1 \\
 0.50, $1'$ & 13 & 0.21 & 0.94 & 0.080 & 4.9 && 188  & 0.092 & 0.25 & 0.021  & 1.1  &&  330 & 0.056 & 0.17 & 0.018 &  1.0 \\
 0.75, $1'$ & 17 & 0.18 & 0.83 & 0.073 & 4.3 && 206  & 0.087 & 0.23 & 0.021  & 1.0  &&  347 & 0.055 & 0.17 & 0.017 &  0.9  \\[0.1cm]

 0.25, $2'$ & 8 & 0.27 &  1.2 &  0.10  & 6.3  && 155  &  0.11  & 0.29  &  0.022 & 1.3  && 295  & 0.059  & 0.18  & 0.018 & 1.1  \\
 0.50, $2'$ & 12 & 0.22 & 0.98 & 0.086  & 5.1 && 182 & 0.095 & 0.26 & 0.022  & 1.1  && 324 &  0.057 & 0.17 & 0.018 & 1.0  \\
 0.75, $2'$ & 15 & 0.19 & 0.86 & 0.078  & 4.6 && 198 & 0.090 & 0.24  & 0.021 & 1.0  && 340  &  0.055 & 0.17 & 0.018 &  0.9 \\[0.1cm]
  
 0.25, $3'$ & 7 & 0.29 & 1.3  & 0.11  & 6.7  && 150  &  0.11  &  0.30 & 0.023  &  1.3 && 290 & 0.060 & 0.19  & 0.019  &  1.1 \\
 0.50, $3'$ & 10 & 0.23 & 1.0  & 0.096  & 5.4 && 174 &  0.098 & 0.26  & 0.022  & 1.1  && 316 & 0.057 & 0.18 &  0.018 &  1.0 \\
 0.75, $3'$ & 13 & 0.21 & 0.92 & 0.087 & 4.8  && 188 &  0.093 & 0.25  & 0.021  & 1.0  && 331 & 0.056 & 0.17 &  0.018 &  0.9 \\[0.1cm]
 
 0.25, $4'$& 6 & 0.32 & 1.4  & 0.13  &  7.0 && 144 &  0.11  & 0.30  & 0.023  & 1.3  && 286  & 0.060 & 0.19 & 0.019 & 1.1   \\
 0.50, $4'$& 8 & 0.25 & 1.1  & 0.11  & 5.7  && 165  &  0.10  & 0.27 & 0.022  & 1.1  && 309 & 0.058 & 0.18 & 0.018 & 1.0  \\
 0.75, $4'$& 10 & 0.22 &  0.97 & 0.098 & 5.1 && 177 & 0.097 & 0.26 & 0.022 & 1.0  && 320 & 0.057 & 0.17 & 0.018 &  0.9 \\

$\mathbf{10^5}$ \textbf{detectors} \\
 0.25, $1'$  & 20 & 0.18 & 0.77 & 0.064 & 4.5 && 201 & 0.088 & 0.24 & 0.021 & 1.1 && 339 & 0.055 & 0.17 & 0.017 & 1.0 \\
 0.50, $1'$ &  31 & 0.14 & 0.63 & 0.051 & 3.5 && 262 & 0.074 & 0.20 & 0.019 & 1.0 && 402 & 0.051 & 0.15 & 0.016 & 0.9\\
 0.75, $1'$ & 41 & 0.12 & 0.55 & 0.044 & 3.0 && 312 & 0.067 & 0.18 & 0.018& 0.8 && 455 & 0.048 & 0.14 & 0.016 &  0.8\\[0.1cm]

 0.25, $2'$ & 17 & 0.19 & 0.85 & 0.069 & 4.8 && 191 & 0.091 & 0.25 & 0.021 & 1.2 && 330 & 0.056 & 0.17 & 0.018 & 1.1\\
 0.50, $2'$ & 27 & 0.15 & 0.68 & 0.054 & 3.7 && 249 & 0.077& 0.21 & 0.019 & 1.0 && 390 & 0.052 & 0.15 & 0.017 & 0.9\\
 0.75, $2'$ & 36& 0.13 & 0.58 & 0.047 & 3.1 && 295  & 0.069 & 0.18 & 0.018 & 0.9 && 439 & 0.049& 0.14 & 0.016 & 0.8\\[0.1cm]
  
 0.25, $3'$ & 14 & 0.21& 0.94 & 0.076 & 5.1 && 181 & 0.095 & 0.26 & 0.021 & 1.2 && 321 & 0.057 & 0.17 & 0.018 & 1.1\\
 0.50, $3'$ & 23 & 0.17& 0.73 & 0.060 & 3.9 && 232 & 0.081 & 0.22 & 0.020 & 1.0 && 376 & 0.053 & 0.16 & 0.017 & 0.9\\
 0.75, $3'$ & 30 & 0.14& 0.63 & 0.052 & 3.3 && 274 & 0.073 & 0.19 & 0.019 & 0.9 && 422 & 0.050 & 0.15 & 0.016 & 0.8\\[0.1cm]
 
 0.25, $4'$& 11 & 0.24 & 1.0 & 0.087 & 5.4 && 171 & 0.10 & 0.27 & 0.022& 1.2 && 313 & 0.058 & 0.18 & 0.018 & 1.1 \\
 0.50, $4'$ & 19 & 0.18 & 0.79 & 0.068 & 4.1 && 216 & 0.086 & 0.23 & 0.020& 1.0 && 363 & 0.054 & 0.16 & 0.017 & 0.9\\
 0.75, $4'$ & 25 & 0.15 & 0.67 & 0.059& 3.4 && 253 & 0.078 & 0.20 & 0.019& 0.9 && 404 & 0.052 & 0.15 & 0.016 & 0.9\\

$\mathbf{10^6}$ \textbf{detectors} \\ 
 0.25, $1'$  & 53 & 0.10 & 0.38 & 0.049 & 2.9 && 260 & 0.073 & 0.20 & 0.019 & 1.0 && 398 & 0.051 & 0.15 & 0.017 & 0.9\\
 0.50, $1'$ & 79 & 0.085 & 0.34 & 0.037 & 2.3 && 357 & 0.060 & 0.16 & 0.017& 0.8 && 495 & 0.046 & 0.13 & 0.015 & 0.8 \\
 0.75, $1'$ & 100 & 0.077 & 0.31 & 0.032& 2.0 && 437 & 0.053 & 0.14 & 0.016& 0.7 && 576 & 0.042 & 0.12 & 0.014  & 0.7\\[0.1cm]

 0.25, $2'$ & 43 & 0.11 & 0.44 & 0.053 & 3.1 && 241 & 0.077 & 0.21 & 0.020 & 1.1 && 378 & 0.052 & 0.16 & 0.017 & 1.0\\
 0.50, $2'$ & 64 & 0.096 & 0.39& 0.040 &2.5 && 327 & 0.064 & 0.17 & 0.018 & 0.9 && 465 & 0.047 & 0.14 & 0.015 & 0.8\\
 0.75, $2'$ & 82 & 0.086 & 0.36 & 0.034& 2.2 && 399 & 0.056 & 0.15 & 0.017 & 0.8 && 537  & 0.044 & 0.13 & 0.015 & 0.8\\[0.1cm]
 
 0.25, $3'$ & 32 & 0.13 & 0.54 & 0.058 & 3.6 && 218 & 0.083 & 0.23 & 0.020 & 1.1 && 355 & 0.054 & 0.16 & 0.017 & 1.0\\
 0.50, $3'$ & 49 & 0.11 & 0.46 & 0.044 & 2.9 && 293 & 0.068 & 0.18 & 0.018 & 0.9 && 431& 0.049 & 0.15 & 0.016 & 0.9\\
 0.75, $3'$ & 63 & 0.099& 0.42& 0.038 & 2.5 && 357 & 0.061 & 0.16 & 0.017 & 0.8 && 496 & 0.046 & 0.13 & 0.015 & 0.8\\[0.1cm]
 
 0.25, $4'$ & 24 & 0.16 & 0.65 & 0.065 & 4.0 && 199 & 0.089 & 0.24 & 0.021 & 1.2 && 337 & 0.055 & 0.17 & 0.018 & 1.1\\
 0.50, $4'$ & 37 & 0.13 & 0.55 & 0.050& 3.1 && 264 & 0.074 & 0.20 & 0.019 & 1.0 && 405 & 0.051 & 0.15 & 0.016 & 0.9\\
 0.75, $4'$ & 49 & 0.11 & 0.48 & 0.043& 2.7 && 320 & 0.065 & 0.17 & 0.018 & 0.9 && 462 & 0.048 & 0.14 & 0.015 & 0.8\\
\hline
\end{tabular}
\caption{The constraints for $w_0$, $w_a$, $w_p$, and $\Omega_K$, and the FoM for various experimental setups using CMB (with lensing), and adding external information like DESI BAO and a $1\%$ $H_0$ prior. The first column contains (f$_{sky}$, beam size) parameters. The constraints for $\Omega_K$ are in units of $10^{-3}$.  Same as table \ref{table:DEparams}, except that $M_{\nu}$ is fixed as in the DETF Fisher matrix. }
\label{table:DEparams_fixedMnu}
\end{table*}


\newpage
\twocolumngrid
\bibliographystyle{apsrev}
\bibliography{CMBlensingExpt}

\end{document}